%% file: paper.tex
\documentclass[submission, Phys]{SciPost}
\input{incl_settings.tex} 
\input{incl_shortcuts.tex}

\graphicspath{{./figs/}}

\begin{document}

\begin{center}{\Large \textbf{
Scaling laws for amplitude surrogates
}}\end{center}

\begin{center}
  Henning Bahl\textsuperscript{1}, 
  Victor Bres\'{o}-Pla\textsuperscript{2},
  Anja Butter\textsuperscript{1,3}, and
  Joaqu\'{i}n Iturriza Ramirez\textsuperscript{3}
\end{center}

\begin{center}
{\bf 1} Institut f\"ur Theoretische Physik, Universit\"at Heidelberg, Germany\\
{\bf 2} Department of Physics, Harvard University, 02138 Cambridge, MA, USA \\
{\bf 3} LPNHE, Sorbonne Université, Université Paris Cité, CNRS/IN2P3, Paris, France \\
\end{center}



\section*{Abstract}
         {\bf 
         Scaling laws describing the dependence of neural network performance on the amount of training data, the spent compute, and the network size have emerged across a huge variety of machine learning tasks and datasets. In this work, we systematically investigate these scaling laws in the context of amplitude surrogates for particle physics. We show that the scaling coefficients are connected to the number of external particles of the process. Our results demonstrate that scaling laws are a useful tool to approach desired precision targets.
         }

\vspace{1pt}
\noindent\rule{\textwidth}{1pt}
\tableofcontents\thispagestyle{fancy}

\section{Introduction}
\label{sec:intro}

Deciphering the fundamental laws of Nature demands precise theoretical predictions and experimental measurements. In high-energy physics, even tiny deviations between Standard Model predictions and collider data can provide critical evidence for new physics. The LHC experiments have already generated petabyte-scale, high-dimensional datasets, and this data volume is expected to increase by an order of magnitude with the High-Luminosity LHC (HL-LHC). Consequently, the development of accurate and fast tools for simulation and inference has become a central and technically demanding frontier in particle physics.

Modern machine learning (ML) has become a powerful and transformative approach to tackling these challenges~\cite{Butter:2022rso,Plehn:2022ftl}. ML techniques can be used across the entire simulation pipeline, finding applications for example in accelerating phase-space integration and sampling~\cite{Bothmann:2020ywa,Gao:2020vdv,Gao:2020zvv,Heimel:2022wyj,Bothmann:2023siu,Heimel:2023ngj,Deutschmann:2024lml,Heimel:2024wph,Janssen:2025zke,Bothmann:2025lwg}, approximating scattering amplitudes~\cite{Bishara:2019iwh,Badger:2020uow,Aylett-Bullock:2021hmo,Maitre:2021uaa,Winterhalder:2021ngy,Badger:2022hwf,Janssen:2023ahv,Maitre:2023dqz,Spinner:2024hjm,Brehmer:2024yqw,Breso:2024jlt,Bahl:2024gyt,Spinner:2025prg,Favaro:2025pgz,Bahl:2025xvx,Herrmann:2025nnz,Bothmann:2025lwg,Villadamigo:2025our,Beccatini:2025tpk,Bahl:2026qaf}, generating complete reconstruction level events~\cite{Hashemi:2019fkn,DiSipio:2019imz,Butter:2019cae,Alanazi:2020klf,Butter:2023fov,Butter:2024zbd,Brehmer:2024yqw,Favaro:2025pgz,Bahl:2025ryd}, and enabling fast and accurate detector simulations~\cite{Paganini:2017hrr,Paganini:2017dwg,Erdmann:2018jxd,Belayneh:2019vyx,Buhmann:2020pmy,Krause:2021ilc,ATLAS:2021pzo,Krause:2021wez,Buhmann:2021caf,Chen:2021gdz,Mikuni:2022xry,Cresswell:2022tof,Diefenbacher:2023vsw,Xu:2023xdc,Buhmann:2023bwk,Buckley:2023daw,Hashemi:2023ruu,Diefenbacher:2023flw,Ernst:2023qvn,Hashemi:2023rgo,Favaro:2024rle,Buss:2024orz,Quetant:2024ftg,Krause:2024avx}. In each stage, ML can significantly enhance both the precision and scalability. Also for inferring theory parameters from data, ML methods like SBI~\cite{Brehmer:2018eca,Brehmer:2018kdj,Brehmer:2018hga,Brehmer:2019xox,Chatterjee:2021nms,Chatterjee:2022oco,Schofbeck:2024zjo,Bahl:2024meb,Bahl:2025mib} or unfolding~\cite{Datta:2018mwd,Andreassen:2019cjw,Bellagente:2019uyp, Bellagente:2020piv, Vandegar:2020yvw, Arratia:2021otl, Backes:2022sph, Diefenbacher:2023wec, Huetsch:2024quz, Butter:2025via} have been demonstrated to outperform classical methods.

For all these applications, ensuring the precision of the trained neural network (NN) is of foremost importance. The simplest way to boost performance is by scaling up the training resources, which consists of a combination of training data, network size and compute. However, increasing these resources blindly can be very inefficient, and in some cases even lead to degraded performance. A way to prevent this is by making use of  scaling laws~\cite{NIPS1988_d1f491a4,hestness2017deep,rosenfeld2019constructive,henighan2020scaling}, which predict that the test loss of a neural network over any task will follow a power law w.r.t.\ the data, compute, and NN size over several orders of magnitude. These laws have been observed empirically in a variety of tasks, and they can also be theoretically motivated in extreme regimes such as large data/network limits~\cite{JMLR:v23:20-1111,bahri2024explaining,hestness2017deep} and random feature models~\cite{lee2017deep,lee2019wide,spigler2020asymptotic,mei2022generalization,bordelon2020spectrum,bordelon2024dynamical}. Scaling laws are especially relevant for the fields of language modeling~\cite{2017arXiv171200409H,henighan2020scaling,kaplan2020scaling,sharma2020neural,2022arXiv220315556H,JMLR:v23:20-1111,pearce2024reconciling,muennighoff2023scaling,hagele2024scaling} and computer vision~\cite{2017arXiv171200409H,sharma2020neural,JMLR:v23:20-1111,alabdulmohsin2023getting}, due to their proven utility for the development of large scale models~\cite{bi2024deepseek}. The scaling behavior that is verified in these contexts has also been observed for particle physics tasks, namely jet classification~\cite{Batson:2023ohn} and amplitude surrogates~\cite{Brehmer:2024yqw,Breso:2024jlt,Spinner:2025prg}.

In this paper, we systematically investigate the occurrence of these scaling laws for amplitude surrogates. We study a large variety of practically relevant processes that cover a diverse set of interaction patterns. We also determine the differences in scaling across architectures by considering a simple multilayer perceptron (MLP) versus the more sophisticated Lorentz Local Canonicalization Transformer (LLoCa-Transformer)~\cite{Spinner:2025prg,Favaro:2025pgz}. As a result of our analysis, we demonstrate that scaling laws across different processes can be bounded by considering just the number of external particles involved in the interaction. Furthermore, we present simple estimations for the amount of training data, compute, and NN size that is required to reach a certain precision target for a broad class of amplitude surrogates where numerical noise is negligible.

The paper is structured as follows. In Sec.~\ref{sec:scaling_laws}, we review scaling laws in deep learning. We describe our amplitude surrogate setup in Sec.~\ref{sec:setup}. After focusing on the $q\bar q\to t\bar t H$ process in Sec.~\ref{sec:saturate}, we establish the presence of scaling laws for a variety of processes in Sec.~\ref{sec:other_processes} and point out how the scaling can be predicted based on the process' complexity. We present conclusions in Sec.~\ref{sec:conclusions}.

\section{Scaling laws in deep learning}
\label{sec:scaling_laws}

Universal approximation theorems in the field of machine learning generally state that a sufficiently wide neural network can in principle approximate any continuous function with an arbitrary degree of precision~\cite{cybenko1989approximation, hornik1989multilayer, leshno1993multilayer}. These theorems legitimize neural networks as a valid tool for extracting complex relationships and insights from data. However, despite this theoretical guarantee, reaching very high precision levels with a neural network often requires a large investment of training resources. This involves scaling up the number of data points $D_\text{train}$, the number of network parameters $N$ and the compute $C$, defined as the number of operations performed during training for a fixed network size. In practice, scaling up the training requires a careful balance, since increasing these variables blindly can result in inefficient trainings. In the worst case scenario, one of the variables being too small can result in a performance saturation that cannot be overcome with any additional increase on the other parameters. 

A simple solution to prevent this issue is to make use of the empirical scaling laws that parametrize the test loss of a neural network as a set of power laws on the variables $N$, $D_\text{train}$ and $C$. These laws represent a valuable tool to predict the training budget that is needed to reach any performance level and also characterize the conditions under which saturation can happen. The general expression for the scaling laws is:
\begin{equation}
    L(X,Y,Z) = (X_c/X)^{\alpha_X} + K_X(Y,Z),
\end{equation}
where $(X,Y,Z)$ denotes any ordering of the variables $(N,D_\text{train},C)$\footnote{We remark that this expression differs from others in the literature where a single expression is used to parametrize the scaling in two or more variables~\cite{bordelon2024dynamical,pearce2024reconciling,muennighoff2023scaling,su2024unraveling,brehmer2024does}. In this paper we choose to focus on 1D scaling laws exclusively and leave a study on multidimensional scaling for future work.}. The first term describes the power law behavior. The parameters $X_c$ and $\alpha_X$ are positive functions of $Y$ and $Z$ that also depend on the network architecture details and the training dataset structure. The loss will hence decrease as a function of $X$ until it reaches a plateau defined by $K(Y,Z)$, which is in turn a monotonically decreasing function in each of its arguments. $X_c(Y,Z)$, $\alpha_X(Y,Z)$ and $K_X(Y,Z)$ are generally observed to converge to constant values once $Y$ and $Z$ become large enough. This behavior implies that whenever the training is limited by either $Y$ or $Z$, network performance will be saturated at the plateau induced by $K_X(Y,Z)$ no matter how much $X$ is increased. On the flip side, once $X$ becomes the only limiting factor during training, the loss will depend exclusively on $X$ and $K_X(Y,Z)$ will converge to the fundamental lower bound of the loss.

In the case of amplitude surrogates, we theoretically expect to reach very good performance, since the amplitude datasets typically feature very little noise. This implies that
\begin{equation}
    K_X(Y,Z) \xrightarrow{Y,Z\to\infty} \text{const.}\simeq 0.
\end{equation}
The remaining noise enters either due to the limited floating number precision or the numerical evaluation of loop integrals at higher orders and is often negligible.

It is possible to predict the scaling coefficients $\alpha_N$ in the infinite $D_\text{train}$ limit by following the empirically validated theory presented in Refs.~\cite{JMLR:v23:20-1111,pnas.2311878121}. This method is based on the conjecture that trained neural networks map the data to a manifold with intrinsic dimension $d$ that encodes the minimal features that are relevant for the learned task. Under this assumption, any neural network featuring piecewise nonlinearities (such as ReLU) will approximate any function on the manifold as a piecewise linear function $\hat{f}$, with the number of pieces being given by the number of network parameters $N$. If the function has a bounded domain on the $d$-dimensional manifold, the average side length of the pieces $s$ will satisfy the relation
\begin{align}
    s \propto \frac{1}{N^{1/d}}\;.
\end{align}
If the function $f$ that is learned by the neural network is also Lipschitz continuous and we assume that the neural network learns the linear dependence of the function up to a negligible error within each piece, then the deviation $|f(x)-\hat{f}(x)|$ will be dominated by quadratic contributions. The upper bound on the deviation is hence proportional to $s^2$, translating into an upper bound on an MSE loss given by:
\begin{align}
    L = \int d^dx |\hat{f}(x) - f(x)|^2\propto s^4 \propto N^{-4/d} \quad\Rightarrow\quad \alpha_N \simeq \frac{4}{d}\;, \label{eq:aN_pred}
\end{align}
The same deduction applies for cross-entropy and KL-divergence losses. In cases where the target $f(x)$ is simpler than a generic Lipschitz function, the loss may be optimized faster and the $\alpha_N$ estimate thus becomes a lower bound, i.e., $\alpha_N \gtrsim 4/d$. Therefore, we can bound the scaling behaviour of the surrogate with the network size as a function of the intrinsic dimension $d$.  

A similar conjecture can be formulated when considering the scaling with training dataset size. If we assume the training dataset to be noise free and given a sufficiently large NN size, the linear piecewise structure of the network will fit a linear function to the environment of each of the training points. If the points are uniformly distributed inside a data manifold with intrinsic dimension $d$, then the average domain size for each of the linear functions will be equal to the average distance between points $r$ and it will scale with the dataset size $D_\text{train}$ as
\begin{align}
    r \propto \frac{1}{D_\text{train}^{1/d}}\;.
\end{align}
Then, under the same Lipschitz continuity conditions as before, we can write down the scaling of the mean-squared error loss as
\begin{align}
    L = \int d^dx |f(x) - f_\text{true}(x)|^2\propto r^4 \propto D_\text{train}^{-4/d} \quad\Rightarrow\quad \alpha_D \gtrsim \frac{4}{d}\;. \label{eq:aD_pred}
\end{align}
In practice, we expect that this estimate will always underestimate the actual scaling exponent more than the $\alpha_N$ estimation. The reason is that a generic neural network will often be able to include non-linear effects around each data point environment. This is empirically observed in Refs.~\cite{rosenfeld2019constructive,kaplan2020scaling}. We also remark that train and test datasets are rarely constructed by drawing i.i.d. from the underlying data manifold, but this does not seem to compromise the validity of the lower bound in practice, as we will demonstrate below. 
A mathematical derivation for classical models connecting the scaling coefficient to the smoothness of the target function can be found in Ref.~\cite{2020arXiv201214501D}.

In the case of the compute $C$, we can hypothesize the dependence of its scaling with $d$ by considering how it can be quantified as a function of $N$ and $D_\text{train}$. As long as the network operations are dominated by the linear layers, the compute $C(N,D_\text{train})$ spent during a network training can be approximated as~\cite{hoffmann2022training,kaplan2020scaling}
\begin{equation}
    C(N,D_\text{train}) \approx \zeta \cdot N_{\rm epochs} \cdot N\cdot D_\text{train},
\end{equation}
where $N_{\rm epochs}$ is the number of epochs and the constant $\zeta$ depends on the architecture details. This formula measures compute in terms of floating point operations (FLOPs). Moreover, we can assume that for a given $(N,D_\text{train})$ there will always exist a compute value above which the loss plateaus, which we call $C_\text{thres}$.
If the spent compute $C$ is smaller than $C_\text{thres}$, the effectively exploited NN size and training dataset size is smaller than their actual values. Since a smaller $N$ or $D_\text{train}$ would increase the loss according to the power laws in Eqs.~\eqref{eq:aN_pred} and \eqref{eq:aD_pred}, we can also expect that
\begin{align}
    \alpha_C \gtrsim \frac{4}{d}. \label{eq:aC_pred}
\end{align}
Again, we emphasize that this is a very rough estimate, since our argument does not take into account how the ratio between $N$ and $D_\text{train}$ for a fixed compute budget or the training batch size can affect performance~\cite{hoffmann2022training,kaplan2020scaling}. In order to mitigate its dependence on $N$ and $D_\text{train}$, a common approach is to optimize their values for each compute budget $C$. These compute-optimal variables have also been observed to have power law relationships with $C$~\cite{hoffmann2022training,alabdulmohsin2023getting,porian2024resolving,hagele2024scaling}. In our case, we will study the scaling as a function of $C$ for hand-picked combinations of $N$ and $D_\text{train}$. As for the batch size, we observe in practice that the scaling for our studied tasks does not depend on a range of batch sizes from $2^6$ to $2^{16}$.

This characterization for the scaling laws is very useful for amplitude surrogates, since the intrinsic dimension will always match the number of degrees of freedom of the interaction amplitude. This is clear if we consider the definition of the intrinsic dimension as the minimum number of variables needed to parametrize a local environment inside the data manifold. Thus, the intrinsic dimension can be trivially computed for each of the studied partonic processes as a function of the number of particles in the final state $n_f$ through the equation $d=3n_f-4$, provided that all particles in the process are on-shell. We will explicitly check these predictions in our numerical investigations below.

Additionally, knowledge of the intrinsic dimension of the data manifold also offers the possibility of assessing the quality of neural network predictions. Several works have demonstrated that the relation between the dimension of the hidden layer representations in trained networks and the dimension of the target space manifold can be used as a predictor for classifier accuracy~\cite{ansuini2019intrinsic} and generalization capabilities~\cite{andriopoulos2025geometric}. Inspired by these findings, we study how to extract the dimension of the representations learned by our neural networks and analyze its relation to the data manifold. To do so, we will use the twoNN method~\cite{NIPS2004_74934548,2017NatSR...712140F,2019arXiv190512784A}, which we introduce in detail in App.~\ref{app:intrinsic_dimension}. As we demonstrate below, we find that the dimension of the learned representation in the last hidden layer by the amplitude surrogates matches the number of degrees of freedom in our datasets. 

Beyond this approach based on the data manifold, there are other theoretical studies in the literature that present insights on the training dynamics behind scaling laws~\cite{lee2017deep,lee2019wide,spigler2020asymptotic,mei2022generalization,bordelon2020spectrum,bordelon2024dynamical}. However, most of these analyses are rooted on random feature or linearized models. These models consist of neural networks where the weights are frozen at initialization and thus restrict training to the learning of linear combinations of the features built randomly by the network. Even though this simplified training regime is known to reproduce the training dynamics of ordinary neural networks at the infinite width limit, it is unable to capture the details of complex feature learning, thus limiting its utility for practical applications.

\section{Computational setup}
\label{sec:setup}

An amplitude surrogate is defined as a neural network that predicts the squared amplitude of a physical interaction as a function of phase-space points. They are trained through a standard regression task where the targets will be the true amplitudes without any noise beyond the numerical accuracy. Our goal is to extract the scaling laws for these networks in each of the variables $N$, $D_\text{train}$, and $C$. To do so, our experimental setup needs to specify both the network architecture and the training strategy. 

\subsection{Network architectures}

We study two architectures: a multi-layer perceptron that uses the momentum invariant pairs of the particle's four-vectors as inputs (MLP-I) and the Lorentz Local Canonicalization (LLoCa) Transformer~\cite{Spinner:2025prg,Favaro:2025pgz}, which achieves Lorentz equivariance by transforming each particle's four vector into a local frame.

Both of these networks are Lorentz-invariant by construction, which has been proven to be an important property to maximise surrogate precision. The main difference between the two is that the LLoCa-Transformer is also equivariant with respect to particle permutations. This property is especially useful when studying processes with several identical external particles~\cite{Brehmer:2024yqw,Spinner:2024hjm,Spinner:2025prg,Favaro:2025pgz}. On the flip-side, the learned frame transformations necessary for the LLoCa-Transformer can impact training stability.

\subsubsection*{MLP-I}

The MLP-I takes as input all the possible bilinear Lorentz invariants of the particles involved in the process,
\begin{align}
    z_{ij} = \log\left(p_i\cdot p_j\right)\;,
\end{align}
where $p_i$ and $p_j$ are the four-vectors of the external particles. We take their logarithms to minimize the effect of large fluctuations across phase space and then apply standardization. We apply this same preprocessing to the amplitude targets.

\subsubsection*{Lorentz Local Canonicalization Transformer}

The four-momentum $p_i$ of each particle $i$ is defined in a global reference frame --- typically the lab frame. LLoCa~\cite{2024arXiv240515389L,Spinner:2025prg,Favaro:2025pgz} transforms each of the four-momenta to its own local reference frame via a learned Lorentz transformation $L_i \in \mathbb{R}^{4\times 4}$,
\begin{align}
    p_i \rightarrow p_{i,L} = L_i p_i\;,
\end{align}
such that $p_L$ is invariant under a general Lorentz transformation $\Lambda$,
\begin{align}
    p_{i,L} \overset{\Lambda}{\longrightarrow} p_{i,L}'=p_{i,L}\;.
\end{align}
This is achieved by constructing $L_i$ such that
\begin{align}
    L_i \overset{\Lambda}{\longrightarrow} L_i' = L_i \Lambda^{-1}\;.
\end{align}
These transformations into the local frames are constructed via a small Lorentz-equivariant network called the frames network. This network predicts a set of three four-vectors for each particle. These four-vectors are orthonormalized and then used to construct the Lorentz transformations $L_i$, which push the four-momenta to their respective local frames.

The LLoCa-Transformer model then operates on these transformed four-vectors by using a standard transformer backbone without ever breaking the Lorentz symmetry. For the self-attention mechanism, the features of particle $j$ have to be transformed into the frame of particle $i$ via the transformation $L_i L_j^{-1}$ if particle $i$ attends to particle $j$. 
No explicit preprocessing is applied to the inputs in order to not break Lorentz equivariance. However, we apply the same preprocessing to the target amplitudes as with the MLP-I.

\begin{table}
    \centering
    \begin{tabular}{l  c c c }
        \toprule
         & abbreviation & MLP-I & LLoCa-Transformer\\
         \midrule
        Network size & $N$ & $\left[10^2, 10^6\right]$ & $6\times 10^6$  \\
        Dataset size & $D_\text{train}$ & $\left[10, 10^6\right]$ & $\left[10, 10^6\right]$ \\
        FLOPs & $C$ & $\left[10^{10}, 10^{17}\right]$ & $\left[10^{10}, 10^{17}\right]$ \\
        \bottomrule
    \end{tabular}
    \caption{Ranges for the number of network parameters, data points, and compute in FLOPs considered for our scaling law tests.}
    \label{tab:ranges}
\end{table}

\subsection{Network training}

\subsubsection*{Scanning ranges for scaling parameters}

In all of our studies, we try to cover the widest range possible in each of the three factors we want to study. The main limitation for this exploration lies in the training time, which we want to restrict to 3 days on an NVIDIA H100 GPU at most for each run to mimic practical training scenarios. This sets a concrete bound on the maximum of both network size and compute ranges. The chosen ranges are listed in Tab.~\ref{tab:ranges}. Our longest runs last 70 GPU hours, within our self-imposed limit.

Whenever we study the scaling law in a given variable, we consider one of the remaining two as a secondary variable and scan over a limited set of representative values. As for the third one, we always fix it to the maximum value allowed by our computing budget. The preferred secondary variables will always be the dataset size, followed by the network size and finally the compute budget.

For all our trainings, we evaluate the networks on a validation dataset consisting of one fourth of the training dataset. Evaluation of the performance on the validation dataset is done every $10^3$ iterations and we do not apply early stopping to ensure that all networks have an equal shot at reaching maximal performance. The test loss is computed on the network that scores the highest on the validation set. 

\subsubsection*{Loss function}

By default, we train both networks on a standard MSE loss. This choice differs from the relative L1 loss that was studied in Ref.~\cite{Breso:2024jlt}, but it matches the loss that is used for almost all amplitude surrogate losses in the literature (see e.g.~Refs.~\cite{Spinner:2024hjm,Favaro:2025pgz,Brehmer:2024yqw}). The conclusions that we obtain with the MSE loss can be directly extrapolated to a setting with the relative L1 loss, since both choices produce the same saturation behavior and the properties of the scaling laws are the same in both scenarios. We verified this for the $q\bar q\to t\bar tH$ amplitude, as shown in App.~\ref{app:L1_vs_MSE}. 

\subsubsection*{Probabilistic amplitude regression}

Instead of seeing amplitude surrogates as a pure regression task, we can also approach it from a probabilistic perspective, allowing us to not only derive a mean prediction $\overline{A}(x)$ but also an associated expected standard deviation $\sigma(x)$. Following Refs.~\cite{Bahl:2024gyt,Bahl:2025xvx}, we can describe the amplitude prediction as a probability distribution $p(A|x)$ for a given phase space point $x$. 

To regress on $p(A|x)$, we minimize the negative log-likelihood over the training dataset $D_\text{train}$,
\begin{align}
    \loss = - \XLangle \log p(A | x) \XRangle_{x \sim D_\text{train}} \;.
    \label{eq:nll}
\end{align}
One useful ansatz for amplitude regression~\cite{Bahl:2024gyt} is to model $p(A | x)$ as a Gaussian distribution,
\begin{align}
    p(A | x) = \normal(A|\,\overline{A}(x), \sigma^2(x)) \;.
\end{align}
Note that this approach can easily be extended to a multi-modal Gaussian to incorporate non-Gaussian uncertainties if needed.
Plugging this ansatz into Eq.~\eqref{eq:nll}, we obtain the heteroscedastic loss
\begin{align}
    \loss_\text{het} = \mean{
        \frac{(A_\text{true}(x) - \overline{A}(x))^2}{2\sigma^2(x)} + \log \sigma(x)
    }_{x \sim D_\text{train}} \;.
    \label{eq:het_loss}
\end{align}
In practice, this means that for each phase-space point, the neural network predicts two outputs: the mean $\overline{A}(x)$ and the standard deviation $\sigma(x)$. For a phase-space point with a large deviation from the true amplitude $A_{\text{true}}(x)$, the NN can enlarge $\sigma(x)$ to compensate for the large deviation. Enlarging $\sigma(x)$ is, however, penalised by the second term in the loss. Note that if $\sigma$ is assumed to be constant, the MSE loss is recovered. Using the heteroscedastic loss does not only provide an uncertainty estimate, but can also result in an increased precision of the mean predicted amplitude with respect to an MSE loss~\cite{Bahl:2025xvx}.

The learned $\sigma(x) \equiv \sigma_\text{syst}(x)$ encodes a systematic uncertainty which captures noise inherent in the data and also uncertainties due to a limited expressivity of the surrogate. For our present study, we do not estimate separately statistical uncertainties --- originating from a lack of training data ---, which can be assessed using, for example, Bayesian NNs~\cite{bnn_early3,Bollweg:2019skg,Kasieczka:2020vlh,ATLAS:2024rpl,Bahl:2024gyt}, repulsive ensembles~\cite{repulsive_ensembles_ml,ATLAS:2024rpl,Rover:2024pvr,Bahl:2024gyt,Bahl:2025xvx}, or evidential regression~\cite{DBLP:journals/corr/abs-1910-02600,2021arXiv210406135M,Bahl:2025xvx}.

We can check the calibration of the estimated systematic uncertainty by considering the systematic pull
\begin{align}
t_\text{syst}(x) = \frac{A_\text{NN}(x) - A_\text{true}(x)}{\sigma_\text{syst}(x)} \;.
\end{align}
If the uncertainties are well-calibrated and statistical uncertainties are negligible, the distribution of $t_\text{syst}$ should follow a unit Gaussian distribution. This allows us to check the validity of the Gaussian ansatz for the likelihood.

We will examine the scaling of the heteroscedastic loss as part of our studies and compare it with the scaling of the standard MSE loss.

\subsubsection*{Maximal Update Parametrization}

When choosing the hyperparameters for our networks, we have observed that the performance is only strongly sensitive to the choice of the learning rate and regularization scheme. However, having to tune two hyperparameters over a wide range of different network sizes, training dataset sizes, and FLOPs spent for training is already not feasible considering our limited compute budget.

To alleviate this issue, we use the maximal update parametrization ($\mu$P) worked out in Ref.~\cite{yang2020feature}. This method addresses a common issue in large network trainings, where it is observed that activation updates scale with the network layer width, leading to imbalanced weight optimizations during training. 

To understand the parameterization, we write down a generic NN in the form
\begin{align}
        f(x) = W_3\phi\left(W_2\phi(W_1 x+ b_1)+b_2\right)\;,
\end{align}
where $W_1 \in \mathbb{R}^{d_\text{in}\times n}$, $b_1\in \mathbb{R}^n$, $W_2\in \mathbb{R}^{n\times n}$, $b_2\in \mathbb{R}^n$, and $W_3\in \mathbb{R}^{n\times d_\text{out}}$. $d_\text{in}$ is the input dimension, $d_\text{out}$ is the output dimension, and $n$ is the dimension of the hidden layers. The activation function is denoted as $\phi$. This generic NN only has one hidden layer, but the adaptations detailed below can be straightforwardly extended to deeper networks.

Using the standard parameterization (SP) of NN, the magnitude of some of the weight updates during training will depend on $n$. This implies that the learning rate has to be re-optimized whenever $n$ changes. The goal of the $\mu$P parameterization is to ensure that the expected size of all weights does not depend on $n$.

To achieve this, the initialization of the weight and bias parameters is adapted, 
\begin{alignat}{4}
    \text{SP}:  &\qquad W_1\sim \normal(0, \nicefrac{1}{d_\text{in}}), &\quad W_{2} \sim \normal(0,\nicefrac{1}{n}), &\quad W_{3} \sim \normal(0,\nicefrac{1}{n}),  \quad &b_{1,2} = 0\;, \\
    \mu\text{P}:&\qquad W_1\sim \normal(0, \nicefrac{1}{d_\text{in}}), &\quad W_{2} \sim \normal(0,\nicefrac{1}{n}), &\quad W_{3} \sim \normal(0,\nicefrac{1}{n^2}),\quad &b_{1,2} = 0\;.
\end{alignat}
Moreover, the learning rates used by the Adam optimizer~\cite{KingmaB14} have to be scaled,
\begin{align}
    \text{SP}:&\qquad \eta_{W_1} = \eta_{b_1} = \eta_{b_2}  = \eta_{W_2} =  \eta_{W_3} = \eta \;,\\
    \mu\text{P}:&\qquad \eta_{W_1} = \eta_{b_1} = \eta_{b_2} = \eta, \quad \eta_{W_2} = \eta  \, n^{-1}, \quad \eta_{W_3} = \eta\,n^{-1} \;.
\end{align}
This parametrization guarantees that maximal performance will be achieved in the infinite width limit. 

Ref.~\cite{2022arXiv220303466Y} also proved that the optimal learning rate will stay constant under $\mu$P for varying hidden dimension. Additionally, empirical tests have shown that $\mu$P also achieves moderate hyperparameter stability against changes in batch size, learning rate scheduler and network depth for some transformer networks\footnote{Several works have attempted to formally extend $\mu$P to achieve efficient hyperparameter transfer across depth, but there is still no general consensus on the matter~\cite{yang2023tensor,bordelon2023depthwise,Blake2024uPTU,Noci2024SuperCO}.}. This enables an efficient learning rate optimization strategy for our networks, by which we only need to tune the learning rate for a small network and then zero-shot transfer it to any larger network size. For more details, we refer to Ref.~\cite{2022arXiv220303466Y}.

We use the $\mu$P parameterization for the MLP-I model. We observe that it is effective for scaling up both network length and width and the scheduler. Once we find the optimal setup for a small network size, we scale it up while keeping the width to length ratio constant. For the LLoCa-Transformer, we applied $\mu$P by hand following the guidelines for the transformer presented in Ref.~\cite{2022arXiv220303466Y}, but we observe a clear instability when transferring the learning rate to high network sizes. We speculate that this is caused by the local frames sub-network and the modified attention mechanism. 

We do not check the effect of $\mu$P on the batch size because we observe that our results are mostly independent of its value. This strategy generally cannot be applied to transfer regularization parameters across networks, so we do not use it for any of our regularization analyses. However, we remark that performance on settings with low training dataset size can be greatly improved by making use of an L2 term in the loss~\cite{l2regularization}.

\subsubsection*{Learning rate scheduling}

We find that a flexible and effective setup for the learning rate scheduling during training is a Cosine Annealing (CA) scheduler~\cite{loshchilov2017sgdr}. A CA scheduler sets the learning rate $\eta$ at the epoch $T_\text{cur}$ via
\begin{align}
    \eta(T_\text{cur}) = \eta_\text{min} + \frac{1}{2}(\eta_\text{max} - \eta_\text{min})\left(1 + \cos\left(\frac{T_\text{cur}\pi}{T_\text{max}}\right)\right)\;,
\end{align}
where $T_\text{max}$ is the final number of epochs.

This setup is robust against changes in the network size thanks to the $\mu$P implementation, but it can lead to suboptimal results in the compute scaling laws if the value $T_\text{max}$ is large enough to push the loss into the saturation regime. Given that performance will stop improving in the proximity of the plateau, it is important that the learning rate is small enough so that the optimal loss is captured accurately. In order to ensure that this is the case for all of our setups, we optimize the learning rate for the setup with the largest $T_\text{max}$.

A constant learning rate would avoid these issues, but using it leads to significantly worse scaling. This result contrasts with the trend observed in language model scaling studies, where it is found that a constant learning rate results in the same optimal compute scaling as using cosine annealing~\cite{pearce2024reconciling}. We could have also considered the ReduceLROnPlateau scheduler~\cite{Bengio2012}, but we would have needed to optimize it separately for every training time setup. Therefore, we opt for the CA scheduler to test the scaling under realistic conditions.

\section{Case study: \texorpdfstring{$q\bar{q} \to t\bar{t}H$}{qq->ttH}}
\label{sec:saturate}

In order to extend the scope of the scaling laws to the context of amplitude surrogates, we first study the $t\bar{t}H$ production process. Observables linked to this interaction have been identified as an interesting probe for potential new physics effects linked to the Higgs sector. On the theory front, it is mandatory to include higher order corrections in order to meet the experimental precision. However, it is currently impractical to compute amplitudes within numerical integration frameworks beyond the leading one-loop QCD corrections due to the prohibitive cost of two-loop amplitude evaluations. This problem can be remedied through the use of interpolation methods, but traditional techniques struggle to achieve acceptable precision levels due to the relatively high-dimensional phase space present in this process and the limited data availability~\cite{Breso:2024jlt}. These conditions set $t\bar{t}H$ production as a prime target for surrogate methods specialized in achieving a high degree of precision under a very strict training data budget.

Despite the effectiveness of neural networks at small training dataset sizes, it has been clearly observed that the interpolation quality stops improving once the performance hits a certain threshold (see Fig.~17 of Ref.~\cite{Breso:2024jlt}). This phenomenon was not present using classical interpolation techniques and it seemingly undermined the utility of neural network surrogates in contexts where data budget is not a limiting factor. In this Section, we identify the cause of this issue and present several solutions for it through the use of scaling laws. 

We focus on the quark channel of this process ($q\bar{q}\to t\bar{t}H$) at leading order as a representative example of the problematic saturation behavior. The $gg \to t\bar{t} H$ has been studied in detail in Ref.~\cite{Breso:2024jlt}, finding a very similar behavior to that of the $q\bar{q}\to t\bar{t}H$ channel. We train our networks on the same dataset that was used in Ref.~\cite{Breso:2024jlt}, which consists of phase space points as inputs and their corresponding squared amplitudes as targets. It was produced through an unweighting sampling method that reproduces the physical distribution of the process across phase space. As found in Ref.~\cite{Breso:2024jlt}, weighted events can also be used for training with comparable performance. All particles in this process are on shell, and the initial quarks are taken to be massless.

When training our networks, we follow the guidelines presented in Section~\ref{sec:setup} except for target preprocessing. Ref.~\cite{Breso:2024jlt} observed a mild improvement in performance from training with non-preprocessed targets, so we match that setup to ensure a fair comparison.

\subsection{Scaling using MSE loss}

\begin{figure}
    \begin{center}
    \includegraphics[width=0.49\linewidth]{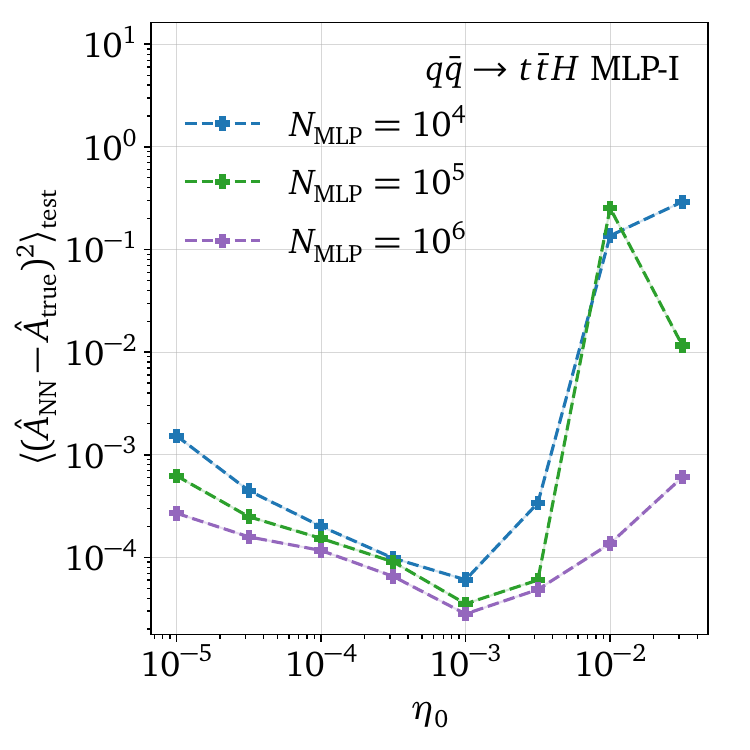}
    \includegraphics[width=0.49\linewidth]{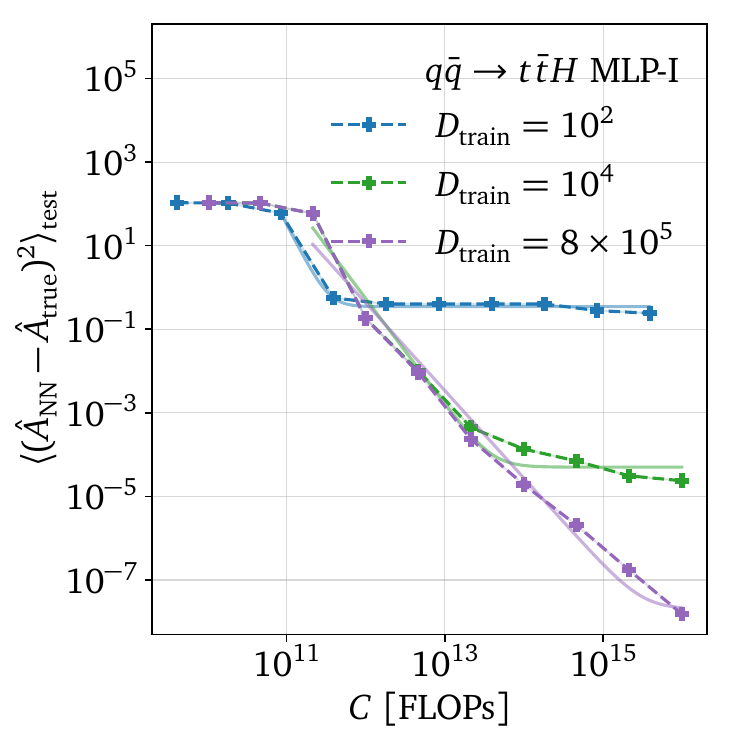}
    \caption{Left: MSE test error for the $q\bar q\to t\bar t H$ amplitude using the MLP-I model as a function of the initial learning rate for different network sizes. Right: Same as left, but the test error is shown as a function of the compute in FLOPs for different training dataset sizes. The fitted power-law functions are depicted as shaded lines.}
    \label{fig:qq_tth_mlp_muP_compute}
    \end{center}
\end{figure}

We first validate the $\mu$P parameterization on our MLP-I networks. In the left panel of Fig.~\ref{fig:qq_tth_mlp_muP_compute} we show as an example the MSE test error as a function of the initial learning rate $\eta_0$ for different network sizes and for a short training. We can clearly see that the test loss is minimal for $\eta_0 \simeq 10^{-3}$ regardless of the network size, confirming that the $\mu$P parameterization works as intended. We use $\eta_0 \simeq 2.7\times10^{-4}$ for the full scaling studies, which we determine by training a small network on our highest training budget. This underestimates the true optimal learning rate on lower $C$ and $D_\text{train}$ values, but we find that the impact on the performance is small.

With this setup, we move on to study the scaling of the test loss as a function of the number of floating point operations (FLOPs) for three fixed dataset sizes and a fixed network size of $10^6$ parameters. To quantify the computing resources, we use FLOPs instead of training time or the number of iterations in order to enable fair comparisons of different network sizes and trainings on different machines. The results displayed in the right panel of Fig.~\ref{fig:qq_tth_mlp_muP_compute} confirm the expected scaling behavior; the test loss decreases as a power law for increasing computing time. We find visible plateaus for $D_\text{train} = 10^2$ and $D_\text{train} = 10^4$. The results obtained here reveal the training dataset size as a very important limiting factor for the network performance that forces the network into the saturation regime even when training the model for a long time. The model benefits from using more than $\sim 10^{14}$ training FLOPs only for $D_\text{train} \simeq 10^6$, which corresponds in this case to $8.6\times10^5$ iterations. For $C \lesssim 10^{11}$, the loss curves have a significantly smaller slope. We identify the slow convergence in this regime as a symptom that the surrogates are overestimating the intrinsic dimension of the data manifold, which we discuss below in more detail.

Moreover, this figure explains the origin of the flat loss in the large data regime observed in Ref.~\cite{Breso:2024jlt}. For the largest dataset size used in the right panel of Fig.~\ref{fig:qq_tth_mlp_muP_compute}, the loss keeps decreasing when more compute is spent. In Ref.~\cite{Breso:2024jlt}, the compute was limited to $10^2$ epochs, corresponding to $\sim 8\times 10^{13}$ FLOPs in that regime, which was not enough to fully exploit the training dataset.

\begin{figure}
    \begin{center}
    \includegraphics[width=0.49\linewidth]{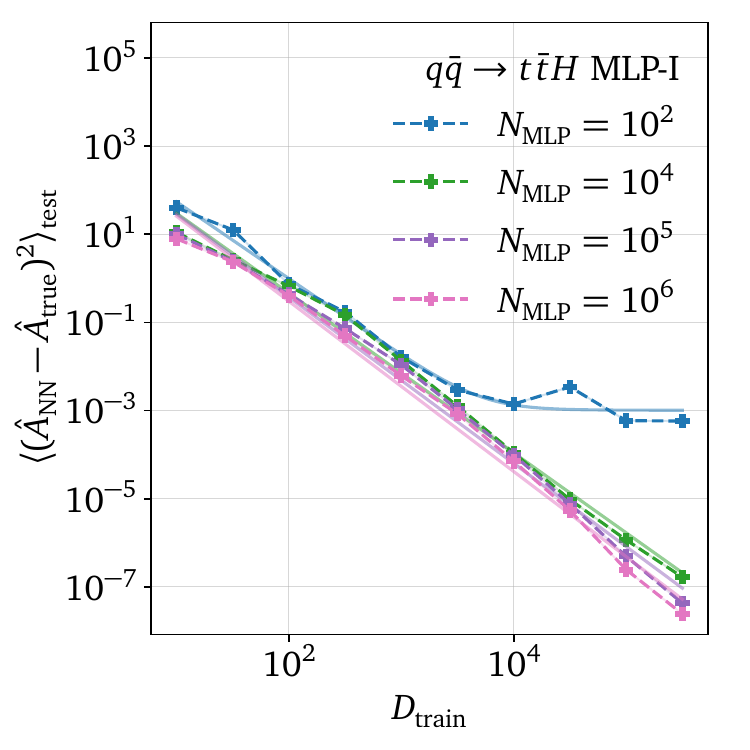}
    \includegraphics[width=0.49\linewidth]{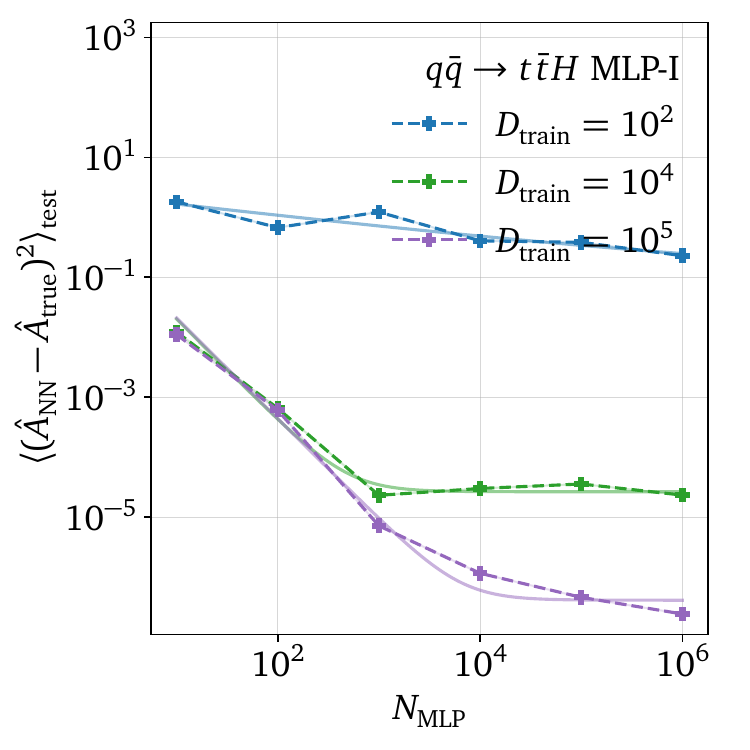}
    \caption{
    Left: MSE test error for the $q\bar q\to t\bar t H$ amplitude using the MLP-I model as a function of the training dataset size for different NN sizes. The fitted power-law functions are depicted as shaded lines. Right: Same as left, but the test error is shown as a function of the NN size for different training dataset sizes.}
    \label{fig:qq_tth_mlp_dtrain_nsize}
    \end{center}
\end{figure}

We now examine performance scaling with the dataset size. Before running the trainings, we need to consider that fixing the FLOPs for a varying dataset size would lead to a reduction of training epochs by a factor $10^6$ from the largest and the smallest dataset. To prevent this, we choose a fixed number of $10^4$ epochs for all our trainings, which was observed to be enough to allow all dataset sizes to reach good performance. We show the results on the left panel of Fig.~\ref{fig:qq_tth_mlp_dtrain_nsize}, where we observe that the test loss follows a power law scaling. There is a visible plateau only for $N_\text{MLP} = 10^2$, confirming the importance of the training dataset size.  Increasing the NN size beyond that only slightly improves the test loss. 

We note here again that the hyperparameters have been optimized for high $C$, $N$, and $D_\text{train}$. While the $\mu$P parameterization guarantees that the learning rate is optimal throughout the considered parameter range, optimizing the weight regularization for low $D_\text{train}$ can further improve performance in this regime. 

Last, we study the scaling of the test loss as a function of the number of network parameters in the right panel of Fig.~\ref{fig:qq_tth_mlp_dtrain_nsize}. The loss curves again feature power law scaling. We also observe that small dataset sizes again play a crucial role on forcing the loss into the saturation regime. Additionally, using a large training dataset significantly improves performance even for a small number of parameters.

\begin{table}
    \centering
    \renewcommand{\arraystretch}{1.3}
    \begin{tabular}{l@{\hskip 15pt} c@{\hskip 15pt} c@{\hskip 15pt} c@{\hskip 15pt} c}
        \toprule
        param.\ $X$ &  2nd param. & $X_c$ & $\alpha_X$ & $K_X$ \\
        \midrule
\hline
\multirow{3}{*}{$C$ [FLOPs]}
 & $D_\text{train}$ = $10^{2}$ & $\left(2.58^{+0.64}_{-0.18}\right)\times 10^{11}$ & $3.67 \pm 0.43$ & $\left(3.47^{+0.54}_{-0.21}\right)\times 10^{-1}$ \\
 & $D_\text{train}$ = $10^{4}$ & $\left(7.85^{+0.55}_{-0.32}\right)\times 10^{11}$ & $2.519 \pm 0.073$ & $\left(50.06 ^{+0.11}_{-3.5}\right) \times 10^{-4}$ \\
 & $D_\text{train}$ = $8.3 \times 10^{5}$ & $\left(6.66^{+0.59}_{-0.31}\right) \times 10^{11}$ & $2.08 \pm 0.032$ & $\left(19.81^{+0.21}_{-1.8}\right)\times 10^{-8}$ \\
 \midrule
                  \hline
\multirow{4}{*}{$D_\text{train}$}
 & $N_\text{MLP}$ = $10^{2}$ & $\left(9.97 ^{+0.78}_{-0.44}\right)\times 10^{1}$ & $1.725 \pm 0.052$ & $\left(7.85 ^{+2.1}_{-0.57}\right)\times 10^{-4}$ \\
 & $N_\text{MLP}$ = $10^{4}$ & $\left(6.51 ^{+0.51}_{-0.29}\right)\times 10^{1}$ & $1.811 \pm 0.027$ & $\simeq 0$ \\
 & $N_\text{MLP}$ = $10^{5}$ & $\left(5.93^{+0.45}_{-0.26 }\right)\times 10^{1}$ & $1.893 \pm 0.027$ & $\simeq 0$ \\
 & $N_\text{MLP}$ = $10^{6}$ & $\left(5.36^{+0.40}_{-0.23}\right)\times 10^{1}$ & $1.928 \pm 0.027$ & $\simeq 0$ \\
\midrule
\multirow{3}{*}{$N$} & $D_\text{train} = 10^2$ & $595\ ^{+360}_{-5.1}$                            & $0.601 \pm 0.040$ & $\left(1.64\ ^{+0.72}_{-0.13}\right) \times 10^{-2}$ \\
                     & $D_\text{train} = 10^4$ & $0.603\ ^{+1.3}_{-0.050}$              & $0.996 \pm 0.040$ & $\left(8.22\ ^{+1.8}_{-0.56}\right) \times 10^{-7}$ \\
                     & $D_\text{train} = 10^5$ & $2.06\ ^{+2.0}_{-0.19}$              & $1.850 \pm 0.070$ & $\left(5.97\ ^{+0.80}_{-0.34}\right) \times 10^{-8}$ \\
                   
        \bottomrule
    \end{tabular}
    \caption{Scaling parameters for $q\bar{q} \to t\bar{t}H$ for MLP-I trained with MSE loss.}
    \label{tab:paramfit_MLP}
\end{table}

Given the observed power-law-like scaling behaviours, we infer the scaling parameters $\alpha_X$, $X_c$ and $K_X$ via a fit in logarithmic space of the MSE between the predicted amplitude and its true value. In Tab.~\ref{tab:paramfit_MLP}, we show the results including uncertainty estimates from the fit, assuming symmetric uncertainties on the logarithm of the parameters. Further details are provided in App.~\ref{app:fits}. The $\alpha_X$ values correspond to the slope in the double logarithmic test loss plots. We find mean values stretching from $\simeq 2 - 3$ for the compute, indicating that the loss decreases by two to three orders of magnitude when increasing the compute by one order of magnitude. Similarly, we find $\alpha_D \simeq 2$ and $\alpha_N \simeq 1-2$.
$K_X$ quantifies the irreducible error originating from factors other than the varied quantity. Here, the fits are relatively uncertain. In particular, our $K_D$ estimates are often unconstrained from below. In such cases, we find that the fitted values of $\alpha_X$ and $X_c$ do not change after repeating the fit without $K_D$, so we approximate $K_D \simeq 0$. 

\begin{figure}
    \begin{center}
    \includegraphics[width=0.49\linewidth]{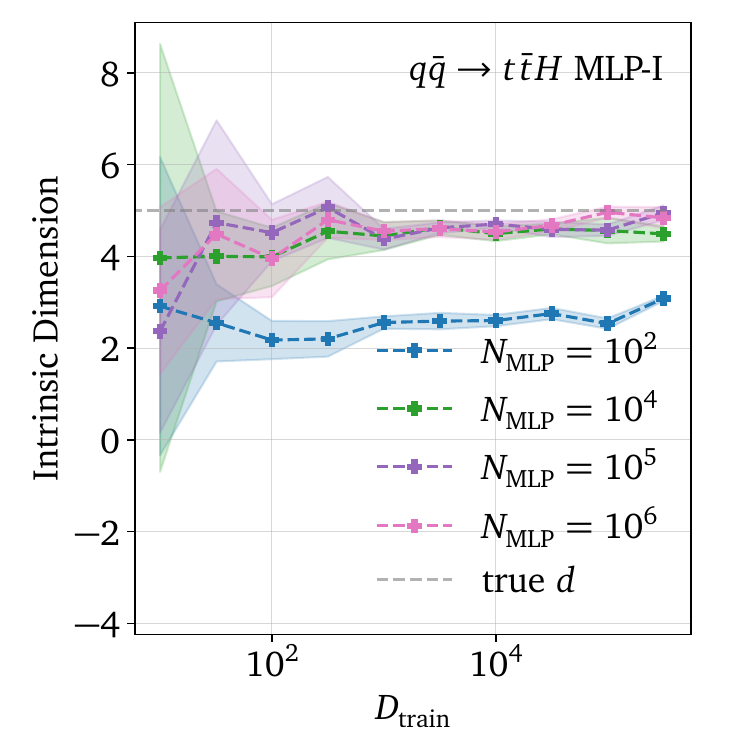}
    \includegraphics[width=0.49\linewidth]{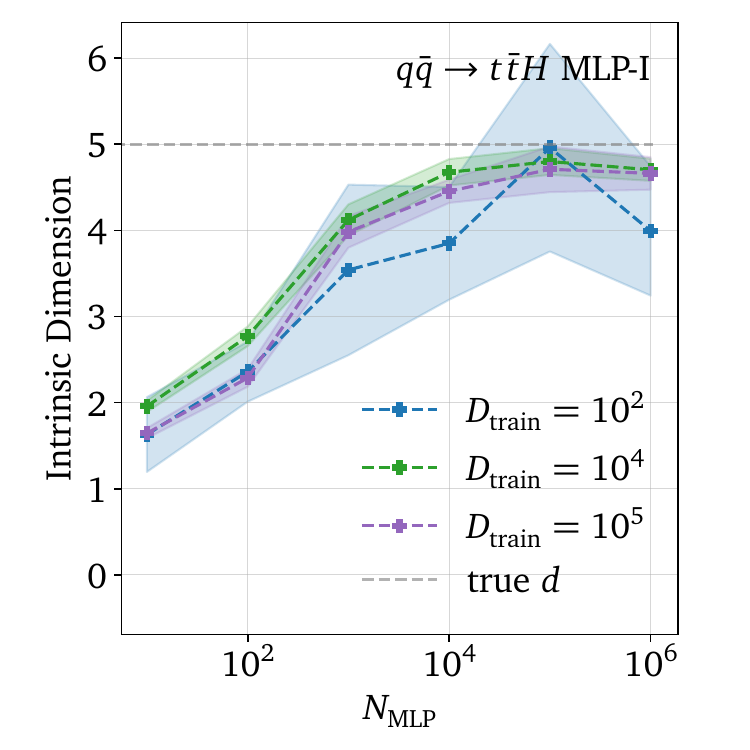}
    \caption{Left: Intrinsic dimension extracted by the $q\bar q\to t\bar tH$ surrogate as a function of the training dataset size for different NN sizes. The color bounds indicate the associated uncertainty. The intrinsic dimension is calculated based on a subset of the final-layer activations with the uncertainties corresponding to the standard deviation obtained by sampling different subsets. Right: Same as left, but the intrinsic dimension is shown as a function of the NN size for different training dataset sizes.}
    \label{fig:intrinsic_dim_ttH}
    \end{center}
\end{figure}

Next, we verify whether the extracted scaling coefficients agree with the theoretical predictions of Eqs.~\eqref{eq:aN_pred}, \eqref{eq:aD_pred}, and~\eqref{eq:aC_pred}. Since $t\bar t H$ production is a $2\to 3$ process, we know that the amplitude depends on five degrees of freedom, implying that the intrinsic dimension is five. Therefore, the lower bound prediction for all $\alpha_X$ is $4/5 = 0.8$, which is fully compatible with most of the results our fits. The only exception is the exponent that we obtain for the scaling law in $N$ with $D_{\rm train}=10^2$, which can be explained by the lack of training data severely limiting the scaling range before saturation. The NNs also correctly extract this intrinsic dimension for $N_\text{MLP}\geq 10^4$ and $D_\text{train} \gtrsim 10^2$, as shown in Fig.~\ref{fig:intrinsic_dim_ttH} and explained in detail in App.~\ref{app:intrinsic_dimension}.

\subsection{Scaling using heteroscedastic loss}
\label{sec:ttH_heteroscedastic} 

After confirming the presence of scaling laws for amplitude regression using a simple MSE loss, we investigate next whether we can also find scaling laws for the more complicated task of predicting the amplitude along with its uncertainty. To do so, we train the MLP-I network on the heteroscedastic loss introduced in Eq.~\eqref{eq:het_loss}.

\begin{figure}
    \begin{center}
    \includegraphics[width=0.49\linewidth]{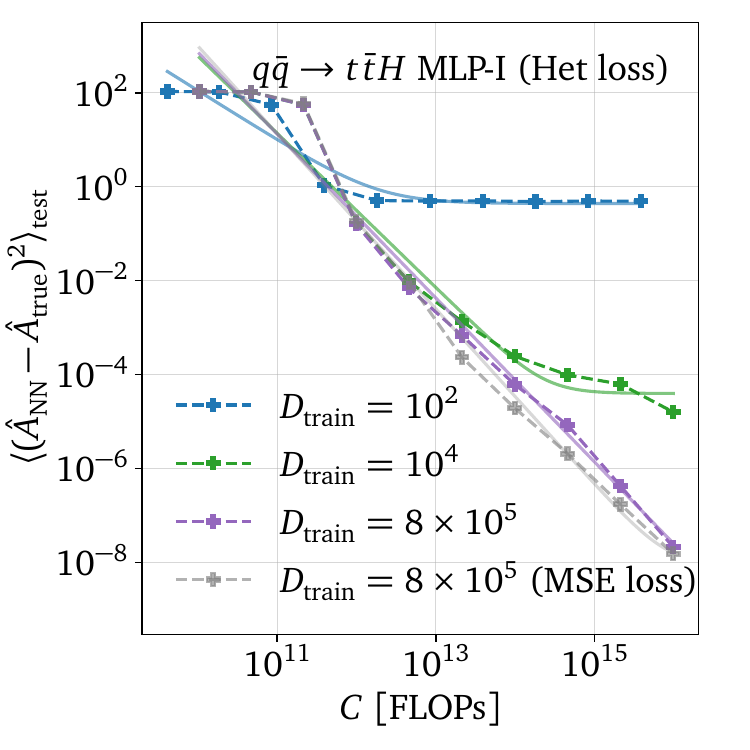}
    \includegraphics[width=0.49\linewidth]{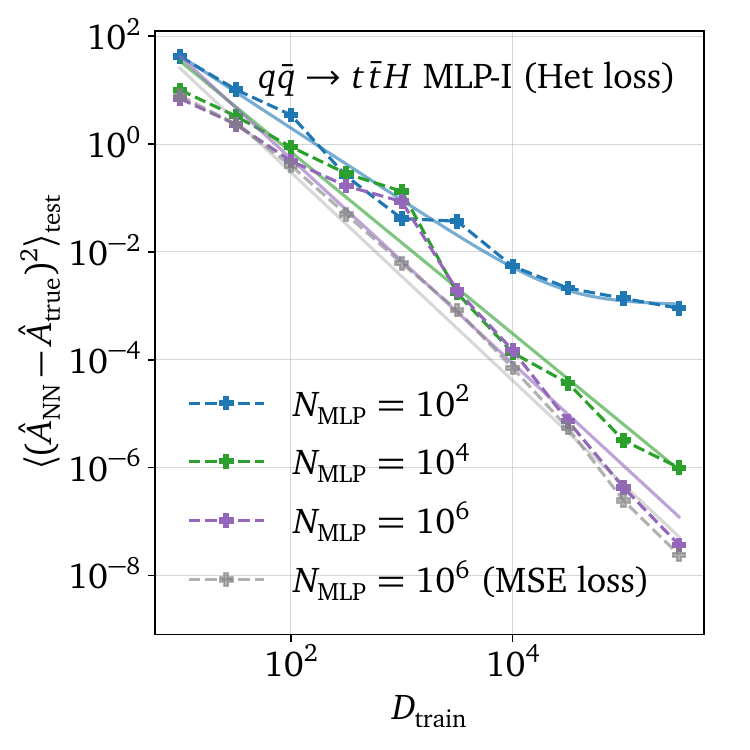}
    \caption{Left: MSE test error for the $q\bar q\to t\bar t H$ amplitude using the MLP-I model trained with a heteroscedastic loss as a function of the compute spent for training in FLOPs for different training dataset sizes. The fitted power-law functions are depicted as shaded lines. Right: Same as left, but the test error is shown as a function of the training dataset size for different NN sizes.}
    \label{fig:qq_tth_mlp_het}
    \end{center}
\end{figure}

The resulting scaling behaviour is shown in Fig.~\ref{fig:qq_tth_mlp_het}. The left panel showing the scaling in $C$ looks almost identical to the right panel of Fig.~\ref{fig:qq_tth_mlp_muP_compute}, and the right panel showing the scaling in $D_\text{train}$ looks similar to the left panel of Fig.~\ref{fig:qq_tth_mlp_dtrain_nsize}. For comparison, we also show the MSE loss curve from these figures for the largest training dataset size (left panel) and the largest NN size (right panel). This demonstrates that using a heteroscedastic loss results in the same scaling behaviour as an ordinary MSE loss.

\begin{figure}
    \begin{center}
    \includegraphics[width=0.49\linewidth]{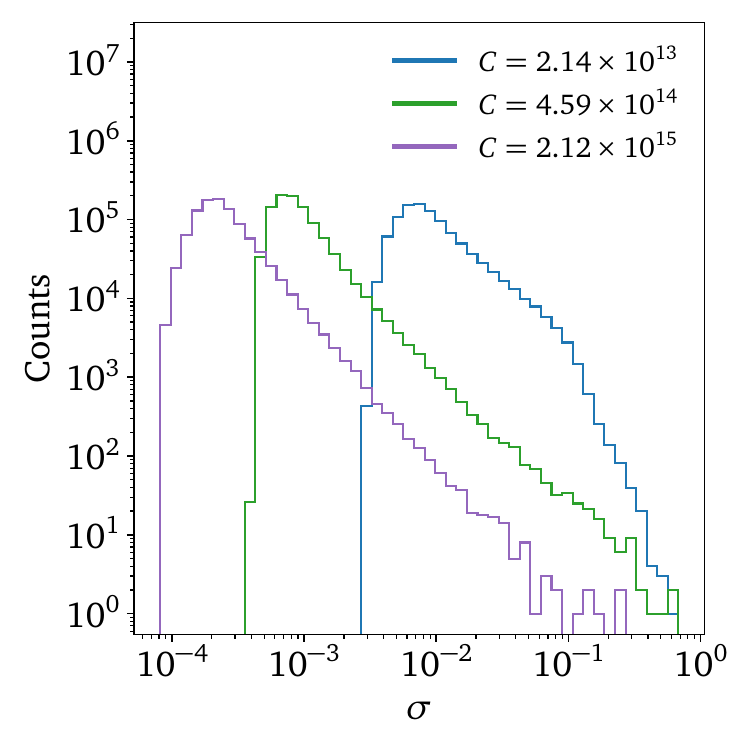}
    \includegraphics[width=0.49\linewidth]{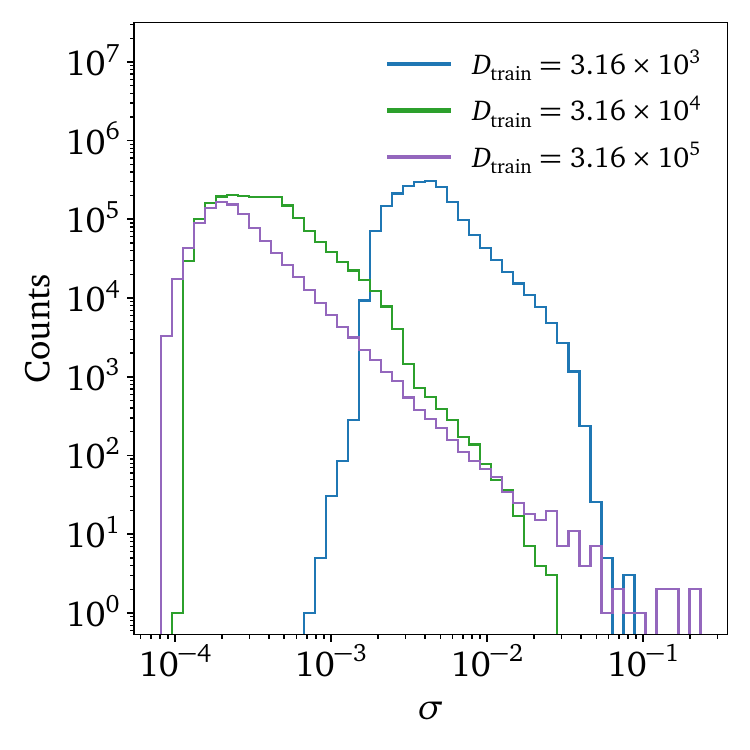}
    \caption{Left: Distribution of learned uncertainty values for the $q\bar q\to t\bar tH$ surrogate for different amounts of spent compute. The dataset size is fixed to $D_\text{train}=3.16\times 10^5$. Right: Same as left, but the distributions are shown for different sizes of the training dataset. The compute is fixed to $C=2.12\times 10^{15}$ FLOPs.}
    \label{fig:qq_tth_mlp_het_sigma}
    \end{center}
\end{figure}

This scaling behaviour is also reflected in the learned uncertainty estimates. We show the distributions of the learned $\sigma$ values in Fig.~\ref{fig:qq_tth_mlp_het_sigma}. In the left panel, we clearly observe that the distribution is shifted towards lower values if the compute is increased. The same pattern is observed if we increase the dataset size, as we show in the right panel. We note that, while the bulk of the $\sigma$ distribution shifts to smaller values as $D_\text{train}$ increases, its high-$\sigma$ tail simultaneously extends further. This reflects the fact that larger training datasets contain more extreme outliers in phase space and thus carry larger uncertainties.

\begin{figure}
    \begin{center}
    \includegraphics[width=0.49\linewidth]{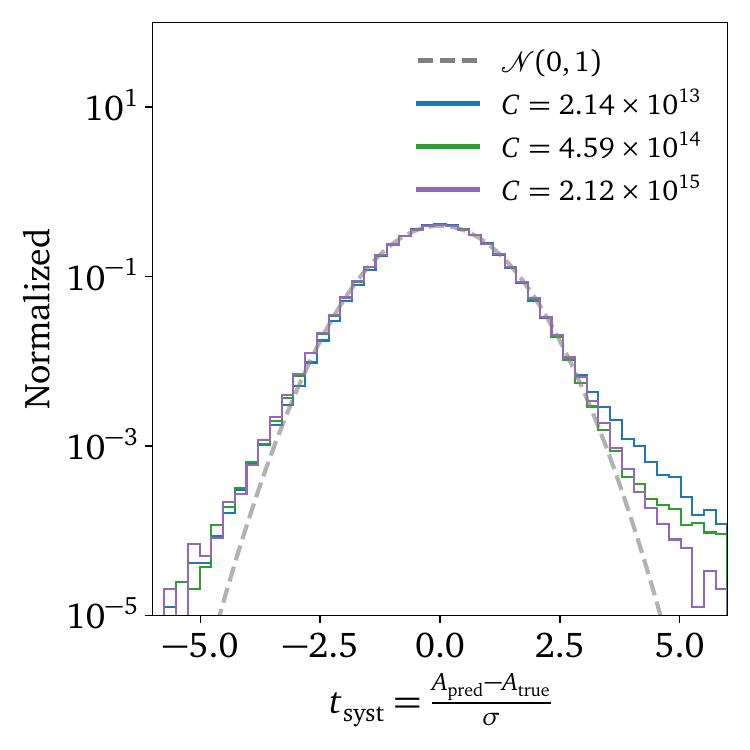}
    \includegraphics[width=0.49\linewidth]{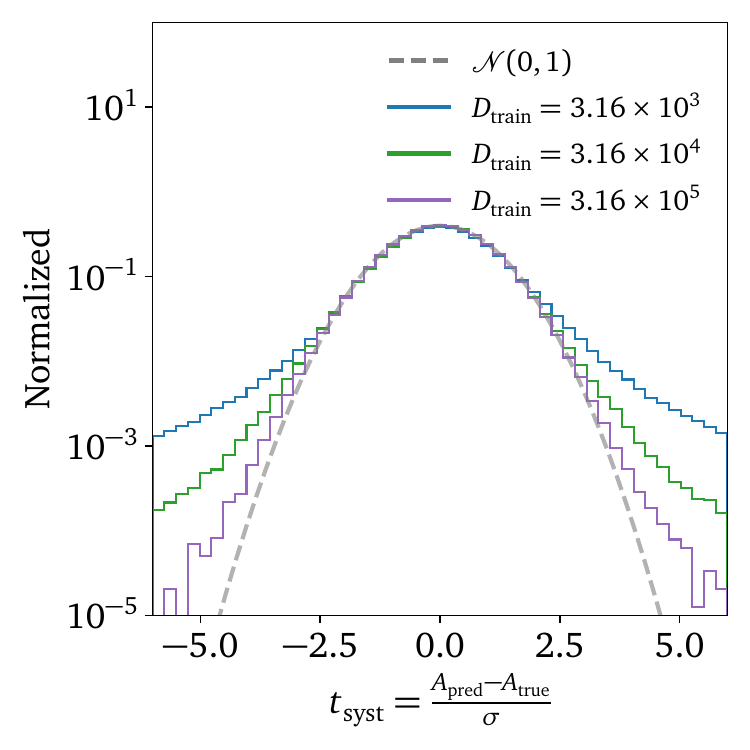}
    \caption{Left: Pull distributions for the $q\bar q\to t\bar tH$ surrogate for different amounts of spent compute. Right: Same as left, but the pull distributions are shown for different sizes of the training dataset.}
    \label{fig:qq_tth_mlp_het_pull}
    \end{center}
\end{figure}

We also investigate the calibration of the learned uncertainty estimate in Fig.~\ref{fig:qq_tth_mlp_het_pull}. The bulk of the distribution is well calibrated following closely a unit Gaussian distribution. We observe a slight overestimation of the uncertainty only at the tails of the distributions. Calibration improves as more compute is spent in the training, especially through increases in the size of the training dataset. For small dataset sizes, different choices for the likelihood form which are able to accommodate for non-Gaussian distributions of the residuals can improve the calibration of the uncertainties, as discussed in detail in Ref.~\cite{Bahl:2026qaf}.

\subsection{Scaling using LLoCa-Transformer}
\label{sec:ttH_lloca}

\begin{figure}
    \begin{center}
    \includegraphics[width=0.49\linewidth]{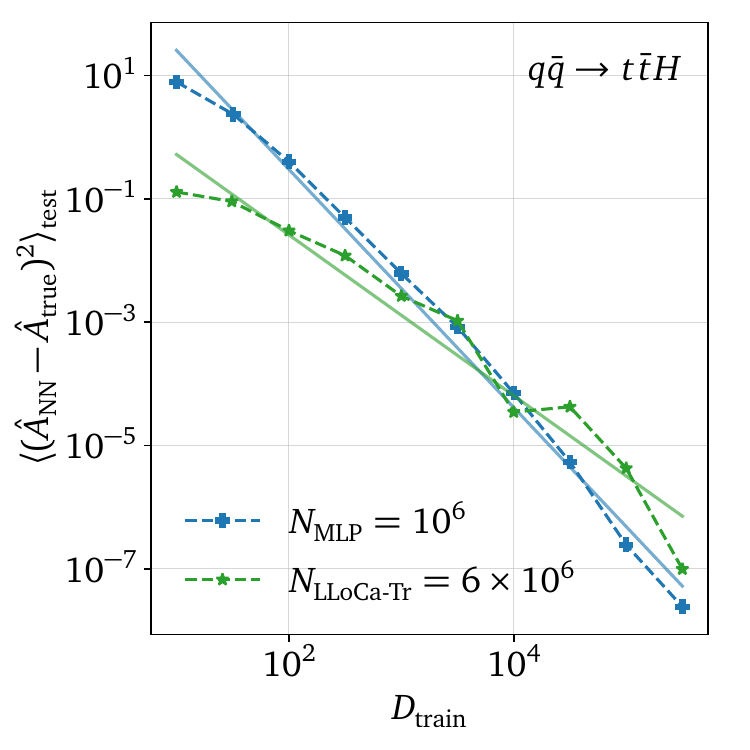}
    \caption{Comparison of the MSE test error for the $q\bar q \to t\bar tH$ amplitude as a function of the training dataset size between the LLoCa-Transformer and the MLP-I model.
    }
    \label{fig:qq_tth_lloca}
    \end{center}
\end{figure}

Finally, we study the scaling behaviour of the LLoCa-Transformer. For all tests with this model, we only examine the scaling law on $D_\text{train}$ because we have observed that the main limitation for the MLP-I performance is the dataset size and we want to check if the scaling in that direction can be improved by deploying a more specialized architecture. We show the comparison between the LLoCa-Transformer and the MLP-I scaling on the $q\bar q \to t\bar t H$ amplitude in Fig.~\ref{fig:qq_tth_lloca}. We observe that the LLoCa-Transformer does not yield a significant advantage over the MLP-I model and it does not appear to follow a power-law relation as neatly as MLP-I. 
We attribute this behavior to a suboptimal choice of hyperparameters for medium and large dataset sizes. In an effort to optimize performance, we performed various grid searches with well educated starting points for each dataset size on the learning rate and L2 regularization parameter. The same grid searches were performed while uncapping the number of epochs and using the maximum amount of computing resources for each run. Tests with different schedulers and warm-up periods were also done, as well as freezing the weights of the local frames sub-part of the network (following the guidelines from Refs.~\cite{Favaro:2025pgz,Spinner:2025prg}). All of these tests only yielded marginal improvements for a few dataset sizes. We attribute this to the general observation that relatively small changes on any hyperparameter often result in large changes in performance, especially on the learning rate. Thus, it is possible that our grid search was too coarse to capture this. As noted above, moreover, the $\mu$P parameterization cannot easily be applied to the LLoCa-Transformer making the hyperparameter optimization significantly more costly in particular for large $D_\text{train}$.

Based on these tests, we conclude that the LLoCa-Transformer surrogate for the $q\bar{q}\to t\bar{t}H$ process is harder to optimize than MLP-I and thus its scaling is less predictable. This observation is in line with the results presented in Refs.~\cite{xie2025tale,elhag2025relaxed} that equivariant constraints can introduce intricate structures into the loss landscape, making their optimization more challenging than that of ordinary neural networks. We speculate further on the cause of this result in Section~\ref{sec:other_processes}, where we also test the LLoCa-Transformer on the $q\bar{q}\to Zgggg$ process.

\section{Scaling laws across processes}
\label{sec:other_processes}

After examining in detail the $q\bar q \to t\bar t H$ amplitude, we now widen the focus and analyze the scaling on a variety of other processes. Our main goals are to study whether the general conclusions we obtained in Sec.~\ref{sec:saturate} can be extrapolated to a wider context, determine the scaling dependence on the intrinsic dimension, and verify whether the postulated behaviour in Eqs.~\eqref{eq:aN_pred}--\eqref{eq:aC_pred} applies for generic amplitude surrogates. We also want to identify if there is any empirical relation between the scaling laws and generic QFT interaction properties. We investigate the following processes:
\begin{itemize}
    \item Jet-associated $Z$ production $(q\bar{q} \to Z+ng, n\in \{1,\ldots,4\})$. This tree-level process has been established as a standard benchmark for testing neural-based amplitude surrogates~\cite{Brehmer:2024yqw,Spinner:2024hjm, Favaro:2025pgz, Spinner:2025prg}. Previous studies have made two key observations: a power-like scaling on the dataset size for the process with $n=4$ gluons, and a clear performance degradation with increasing particle multiplicity, the degree of which being very dependent on the surrogate architecture.
    \item Jet-associated di-photon production $(gg \to \gamma \gamma +ng, n\in \{1,2\})$. This is a loop-induced process which has been studied previously in Refs.~\cite{Aylett-Bullock:2021hmo,Badger:2022hwf,Bahl:2024gyt,Bahl:2025xvx}. It presents the same permutation symmetry pattern as the jet-associated $Z$ boson production with respect to the outgoing gluons in the final state, but it also presents a boson pair initial state and a fully massless particle ensemble. 
    \item Jet-associated $WZ$ production $(q\bar{q} \to WZ+ng, n\in \{0,\ldots,2\})$, and $WWZ$ production $(q\bar{q} \to WWZ)$. These tree-level processes have never been studied in the context of amplitude surrogates and their associated datasets were produced specifically for this study. These amplitudes were chosen with the goal of exploring environments with a large diversity of particle species and masses in the final state. They also represent variations on the jet-associated $Z$ production processes, allowing for an easy comparison with the results from those amplitudes. More specifically, the $WZ$ production allows us to study the gluon multiplicity scaling with an extra heavy boson in the final state, whereas the $Z+nW$ processes include $W$ bosons instead of gluons in the final state. All of these datasets were produced following the same strategy as for the $Z$ production datasets~\cite{Brehmer:2024yqw,Spinner:2024hjm}.    
\end{itemize}
This set of processes allows us to study the surrogate scaling dependence on the following properties: particle multiplicity, particle mass patterns, boson/fermion structures in the initial state, and the degree of permutation invariance when exchanging identical particles. For all processes, we are using unweighted phase-space samples for training, massless initial state particles whose momenta are sampled according to the proton PDFs, and all particles are on-shell. We neglect the effect of the initial state quark flavors; we always consider up quarks in practice for the amplitude computations. The fitted scaling law coefficients are collected in App.~\ref{app:fits}.

\begin{figure}
    \begin{center}
    \includegraphics[width=0.49\linewidth]{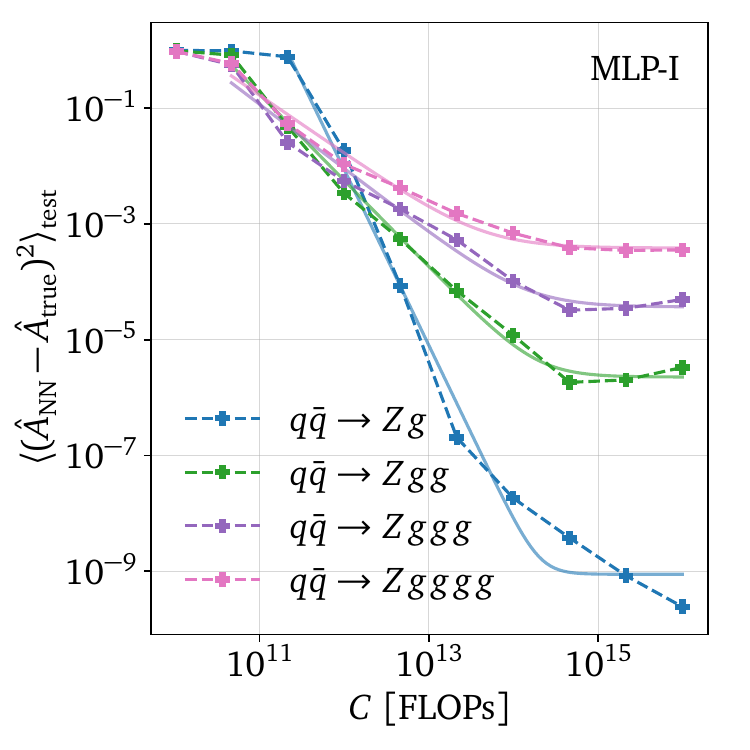}
    \includegraphics[width=0.49\linewidth]{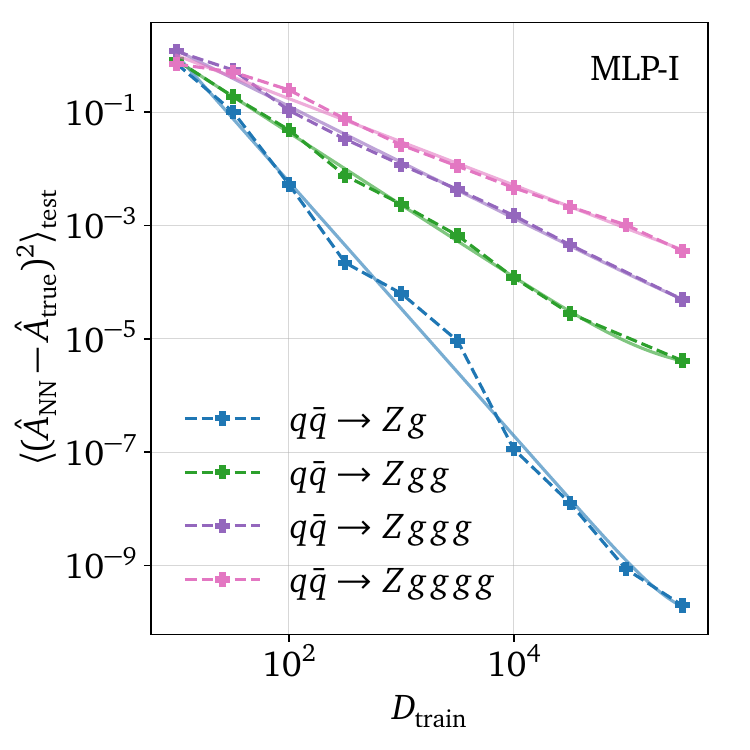}
    \includegraphics[width=0.49\linewidth]{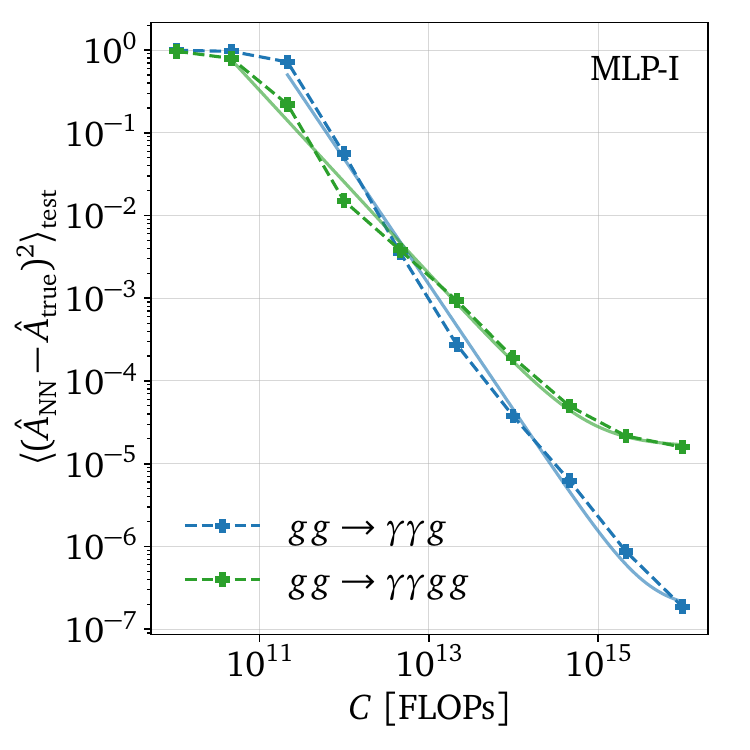}
    \includegraphics[width=0.49\linewidth]{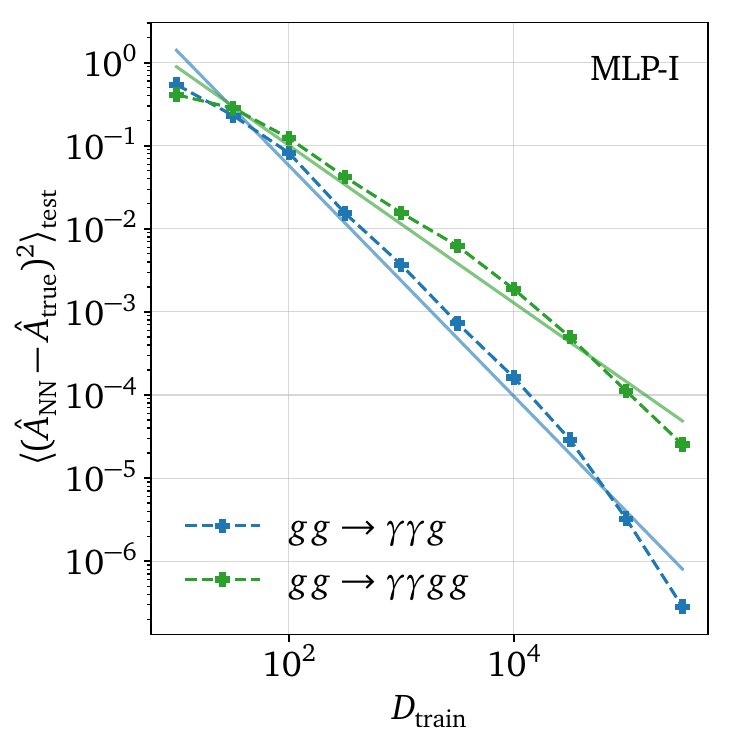}
    \includegraphics[width=0.49\linewidth]{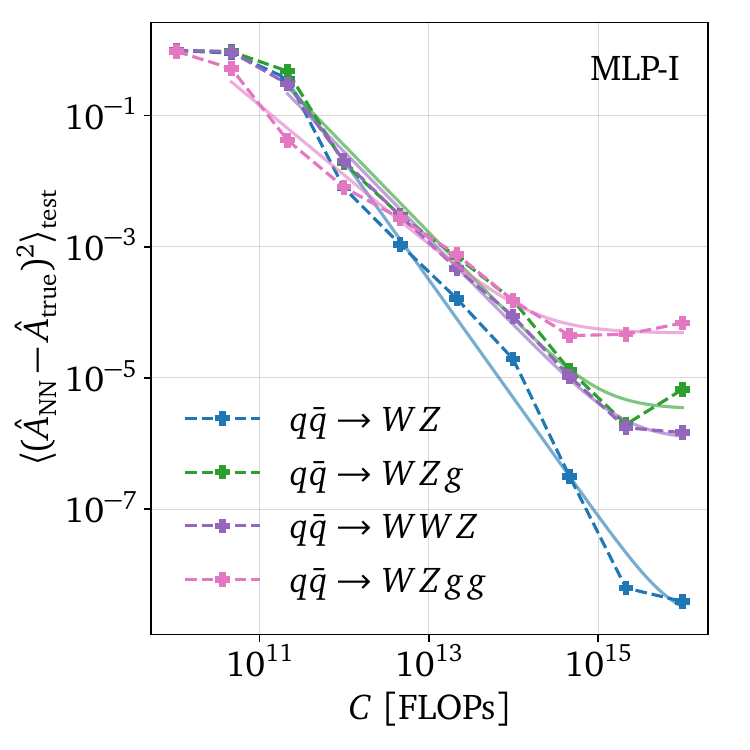}
    \includegraphics[width=0.49\linewidth]{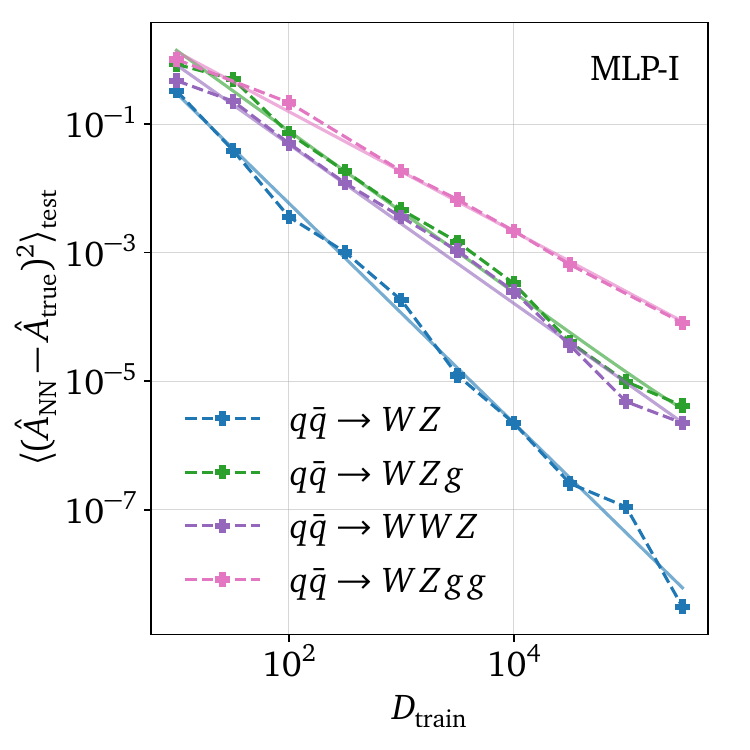}
    \caption{MSE test error as a function of the compute spent for training in FLOPs (left) and the training dataset size (right) for the $q\bar q\to Z + n g$ (upper row), $gg\to \gamma\gamma g, \gamma\gamma g g$ (middle row), and $q\bar q\to WZ,WZg, WWZ, WZgg$ amplitudes (lower row).}
    \label{fig:all}
    \end{center}
\end{figure}

We start by deriving the scaling laws on the MLP-I architecture on all of these processes, with the results shown in Fig.~\ref{fig:all} and Apps.~\ref{app:Zjets},~\ref{app:diphoton} and~\ref{app:wz}. We observe the same qualitative scaling behavior as with the $q\bar{q}\to t\bar{t}H$ interaction. Namely, the dataset size $D_{\rm train}$ is the main performance driver, the scaling in $C$ is always constrained by limited data availability and there is no benefit
from increasing network size past $10^4$ parameters. In the case of previously studied processes, our results validate all past observations from the literature~\cite{Spinner:2024hjm,Breso:2024jlt,Spinner:2025prg,Favaro:2025pgz}. We also clearly see that the slopes of the $C$ and $D_\text{train}$ scaling curves decrease with the number of external particles, as expected from the lower bounds on the scaling coefficients shown in Eqs.~\eqref{eq:aN_pred}--\eqref{eq:aC_pred}.

\begin{figure}
    \begin{center}
     \includegraphics[width=0.49\linewidth]{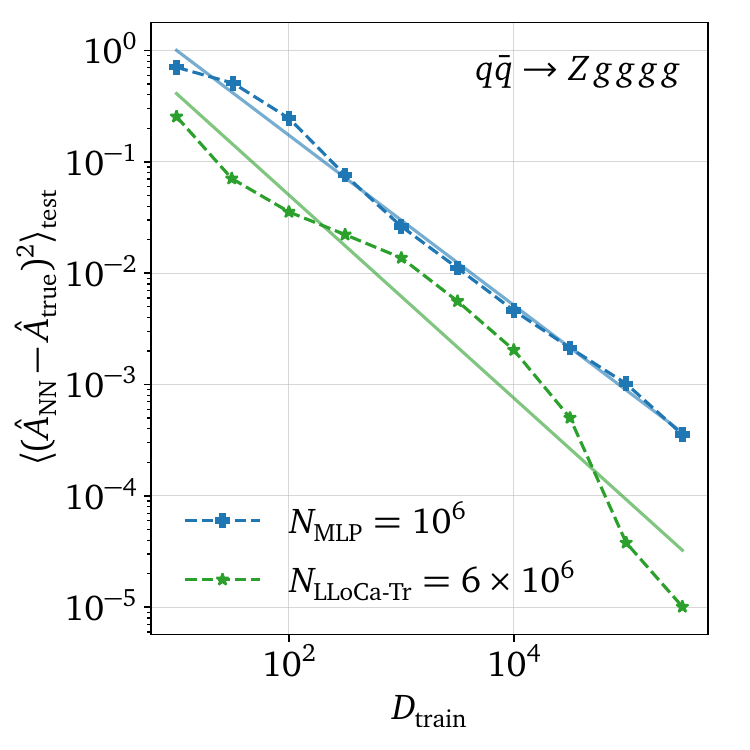}
    \caption{MSE test errors for the $q\bar q\to Zgggg$ amplitude comparing the MLP-I and LLoCa-Transformer models as a function of the size of the training dataset.}
    \label{fig:zgggg_lloca_comparison}
    \end{center}
\end{figure}

We now compare the scaling of MLP-I and LLoCa-Transformer on the $q\bar{q}\to Z+4g$ process in Fig.~\ref{fig:zgggg_lloca_comparison}. We observe the same behaviour for the $\alpha_D$ exponent for both networks with respect to the number of particles in the final state, but $D_{\text{data},c}$ is much smaller for LLoCa-Transformer, resulting in a flat performance improvement for every $D_\text{train}$. This contrasts with our results in Section~\ref{sec:ttH_lloca}, but is fully in line with the observations from previous studies~\cite{Spinner:2025prg,Favaro:2025pgz}. In this case, we also find that the LLoCa-Transformer performance is more stable against hyperparameter changes. One explanation for this may be that the $q\bar{q} \to Z g g g g$ is better adjusted to the symmetry constraints that are enforced by the LLoCa-Transformer, which may help supersede the optimization difficulties. Indeed, this behavior matches the results in Ref.~\cite{elhag2025relaxed}, where it is observed that equivariant neural network trainings becomes simpler as the symmetry constraints become more aligned with the symmetries of the data. However, a more dedicated study is necessary to fully clarify the properties of the LLoCa-Transformer scaling.  

Concerning the estimation of the intrinsic dimension, we observe that all well converged networks are able to recover the true dimension of the processes, as we show in detail in App.~\ref{app:intrinsic_dimension}. In order to achieve this, the networks only require $10^4$ learnable parameters and $10^4$ training data points, with an even smaller training budget being sufficient for some of the processes. We also note that we incur a slight overestimation on some of the datasets with trainings on more than $10^5$ points. We do not have a clear explanation for the cause of this overshooting, but it is mild enough that it does not compromise the final intrinsic dimension estimation. 

\begin{figure}
    \begin{center}
    \includegraphics[width=0.98\linewidth]{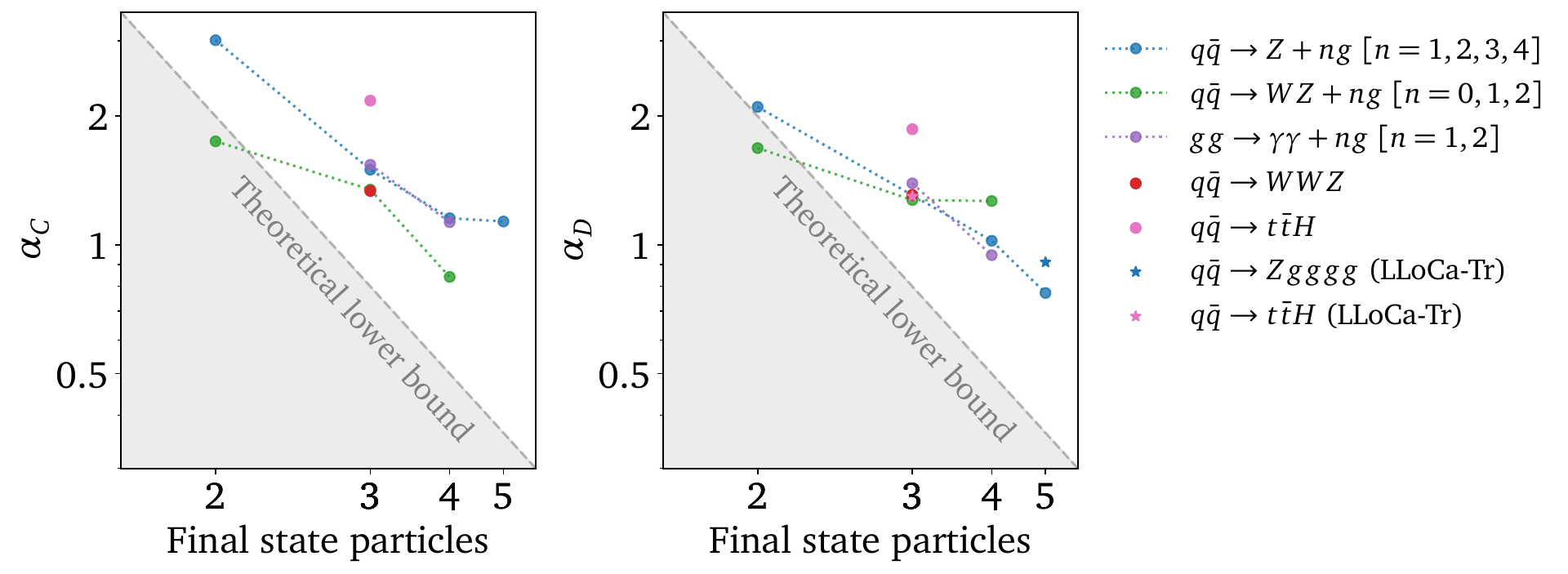}

    \caption{Left: Power law exponent $\alpha_X$ as a function of the number of particles in the final state of each process for the scaling on computing resources. Right: Same as left for the scaling on the dataset size.}
    \label{fig:alphas}
    \end{center}
\end{figure}

We now move on to compare the scaling behavior of the different processes. We plot in Fig.~\ref{fig:alphas} the scaling exponents $\alpha_D$ and $\alpha_C$ as a function of the number of particles in the final state $n_f$, which determines the intrinsic dimension of the process through the relation $d = 3n_f - 4$. The fitted exponents for all processes are in good agreement with the theoretical bound of Eqs.~\eqref{eq:aD_pred} and~\eqref{eq:aC_pred}, shown as dashed lines in Fig.~\ref{fig:alphas}. This result also demonstrates that the scaling on dataset size and compute across processes depends mostly on the total particle multiplicity and not on any details about particle species or symmetry patterns. The choice of architecture has a significant impact on the scaling, but it never compromises the validity of the theoretical bound.  

The only outliers on this trend are the processes with just two particles in the final state. One possible explanation for this behavior is the singularity structure of the $2\to 2$ amplitudes. As discussed also in Ref.~\cite{Bahl:2026qaf}, the precision of the NN surrogates drops close to sharp features of the amplitude caused by the forward/backward scattering singularities, which are regulated by the masses of the final-state particles and minimal transverse momentum cuts. Within these low-precision regions, the NN might not be able to learn the linear dependence of the amplitude up to a negligible error, as assumed in the derivation of Eqs.~\eqref{eq:aN_pred} and~\eqref{eq:aD_pred}. This would effectively lower the scaling coefficient. In the case of the low-dimensional $2\to 2$ processes, more points lie close to these singularities, making this effect more pronounced. We can also compare the singularity structure between the two $2\to 2$ processes we studied by studying how their amplitudes scale with the Mandelstam variables, which completely capture the functional structure of the amplitude. In particular, the $q\bar{q}\to WZ$ amplitudes features singularities that scale as $1/t^2$, $1/u^2$, $1/(st)$, $1/t$ and $1/u$ whereas the $q\bar{q}\to Zg$ process only has poles of the last three types. Thus, we see a clear correlation between the complexity of the amplitude singularities and the value of the neural scaling exponents for a fixed final state multiplicity. These observations reveal a simple method to estimate the resources needed to reach a certain surrogate precision target on a broad variety of physical process. Given that the lower bound on the scaling exponents only depends on the number of particles in the final state, the $X_c$ parameters can be estimated by training a low-cost surrogate --- using a small training dataset, a small NN size, and a low number of FLOPs. Concerning the plateau parameters, the strong dependence of the scaling on the dataset size and the general observation that $K_D \sim 0$ allow us to approximate $K_N$ and $K_C$ as the predicted losses from the $D$ scaling law. Once the scaling laws are estimated, we will have access to an upper bound on the resources needed to reach any target surrogate performance. This upper bound can be further tightened by a second low-cost surrogate --- using slightly more resources --- to derive a better estimate of the scaling factors.

We can demonstrate the usefulness of scaling laws for amplitude surrogates by predicting the dataset size that we would need to reach a target precision of $10^{-3}$ on the $gg\to\gamma\gamma gg$ process in a regime where we are not limited by network size or compute. Rather than relying on the full scan, we use only two inexpensive trainings, at $D_{\rm train}=10^{2}$ and $D_{\rm train}=3.2\times10^{2}$, which together cost less than four minutes of GPU time and yield test losses of $0.123$ and $0.042$. Their ratio fixes the scaling exponent to $\alpha_D=0.94$, in excellent agreement with the value $0.939\pm0.031$ obtained from the full fitting procedure, and anchoring the power law on the second point gives $X_D=10.7$. We can now use these two estimates of the scaling law parameters to predict that a precision of $10^{-3}$ will be achieved at $D_{\rm train}=1.7\times10^{4}$, which matches the dataset size at which the measured test loss actually crosses $10^{-3}$. We stress that the measured exponent is essential here: replacing it with the lower bound $\alpha_D\gtrsim 4/d=0.5$ while anchoring on the same run would instead predict $D_{\rm train}=5.6\times10^{5}$, over-estimating the requirement by a factor of 30. 

We emphasize that this procedure relies on the assumption that numerical noise in the evaluation of the true amplitude is negligible. If noise becomes relevant at a certain level, the accuracy of the surrogate cannot be further improved without further assumptions about the noise. We also remark that our results apply strictly to partonic leading order processes. We expect that our results will apply to hadronic processes, since they can be modeled as simple compositions of partonic subprocesses. We leave the study of more complex processes to future work.

\section{Conclusions}
\label{sec:conclusions}

Fast high-precision Monte-Carlo event generation is a central requirement for fully exploiting the increasingly large datasets collected at the LHC. Surrogate amplitudes are a key ingredient for speeding up event generation~\cite{Janssen:2023ahv,Herrmann:2025nnz,Beccatini:2025tpk,Villadamigo:2025our,Bahl:2026qaf}. For the practical application of surrogate amplitudes it is important to know how many resources --- size of training dataset, size of NN, compute to be spent --- are needed to reach a certain precision target. In the present work, we have demonstrated that scaling laws are a useful tool for answering this question. 

Throughout a variety of applications and fields, the performance of NNs appears to scale like a power law on the size of the training dataset, the size of the NN, and the spent compute. If one of these quantities is not sufficiently large, increasing the other resources will not further improve the performance. We discussed how a lower bound on the scaling coefficients can be set based on the intrinsic dimension of the data. For amplitude surrogates this intrinsic dimension is known and completely fixed by the number of external particles.

As a first step in our empirical studies, we focused on the $q\bar q\to t\bar t H$ process. In the surrogate training for this process, we established the emergence of power-law like scaling for all three relevant quantities. We found the training dataset size to be the most important factor followed by the spent compute. The NN size only played a minor role. These conclusions are strictly valid only inside the training variable ranges that we have studied, but we expect every practical surrogate analysis to follow a training regime within the limits we considered. 
Moreover, we showed that the emergent scaling laws are not only a feature of a specific loss --- i.e., the mean-squared-error loss used for most of our studies --- but also appear for other loss functions like a heteroscedastic loss. In this context, we also studied the calibration of the associated systematic uncertainty finding the expected behaviour across the studied range of resources. Besides different loss functions, we furthermore verified that the scaling law exponents are not peculiar to a certain architecture --- i.e., the mainly employed MLP using Lorentz invariants as input --- but also appear for the LLoCa-Transformer, a Lorentz- and permutation-equivariant transformer architecture. However, we remark that the scope of our analysis on this matter is quite limited, and a more dedicated study should be performed to fully assess the dependence of scaling laws on architecture details.

After establishing the existence of power-law like scaling for a variety of architectures and objectives for the $q\bar q\to t \bar t H$ process, we widened our focus and studied ten other processes differing not only in the number of external particles but also in the nature of the external particles and the involved interactions. For all these processes, we found power-law scaling of the test losses. If the process is (partly) invariant when permuting the external particles, using a permutation-equivariant architecture like LLoCa-Transformer boosts the performance but does not change the scaling coefficients significantly. Moreover, we found empirical evidence for the relation between the scaling law coefficients and the number of external particles, confirming the theoretical expectations. Since the number of external particles and thereby the intrinsic dimension are known before training, this opens up an intriguing avenue to predict an upper bound on necessary resources to reach a certain precision target. Based on a low-cost surrogate --- using a small training dataset, a small NN size, and a low number of FLOPs ---, an anchor point for the scaling curves can be set. The lower bounds on the scaling coefficients then allow one to derive an upper bound for the needed resources to reach any desired precision target. This estimate can be further tightened by a second surrogate using slightly more resources allowing to constrain the true scaling coefficients more tightly.

As a by-product of this study, we have observed that the dimension of the representations learned by our networks closely approximates the number of degrees of freedom of the associated processes. This implies that amplitude surrogates can be used as simple estimators for the intrinsic dimension of the input data manifold. To the best of our knowledge, this is the first time that this association has been validated on datasets where the true dimension is dictated by physical symmetries. This differs from past studies involving teacher-student models, where the predictions are performed on datasets specifically designed with an underlying low-dimensional structure~\cite{JMLR:v23:20-1111,pnas.2311878121}. This hints at the possibility of neural networks being useful tools for probing data spaces featuring a high degree of symmetry structure.  

The observed universality of power-law scaling opens the door to systematic and predictable surrogate training. This will be an essential component for making amplitude surrogates a reliable component for next-generation event generators.
 
\section*{Acknowledgments}

We thank Jonas Spinner for helpful discussions about LLoCa. We thank Bertrand Laforge for insightful discussions on optimization and uncertainty quantification.
AB and JIR gratefully acknowledge the continuous support from LPNHE, CNRS/IN2P3, Sorbonne Université and Université Paris Cité.
The work of AB, HB and VBP is supported by the Deutsche Forschungsgemeinschaft (DFG, German Research Foundation) under grant
396021762 – TRR 257 Particle Physics Phenomenology after the Higgs Discovery.
HB acknowledges support by the state of
Baden-Württemberg through bwHPC and the German Research Foundation (DFG) through
the grant INST 39/1232-1 FUGG. VBP acknowledges financial support from the Real Colegio Complutense at Harvard University Postdoctoral Research Fellowship. The work of JIR is co-funded by the European Union’s Horizon Europe research and innovation programme Cofund SOUND.AI under the Marie Skłodowska-Curie Grant Agreement No 101081674. Views and opinions expressed are however those of the author(s) only and do not necessarily reflect those of the European Union or the granting authority. Neither the European Union nor the granting authority can be held responsible for them.

\clearpage
\appendix

\section{Supplementary results}

\subsection{Scaling with other precision metrics}
\label{app:L1_vs_MSE}

\begin{figure}[H]
    \begin{center}
    \includegraphics[width=0.49\linewidth]{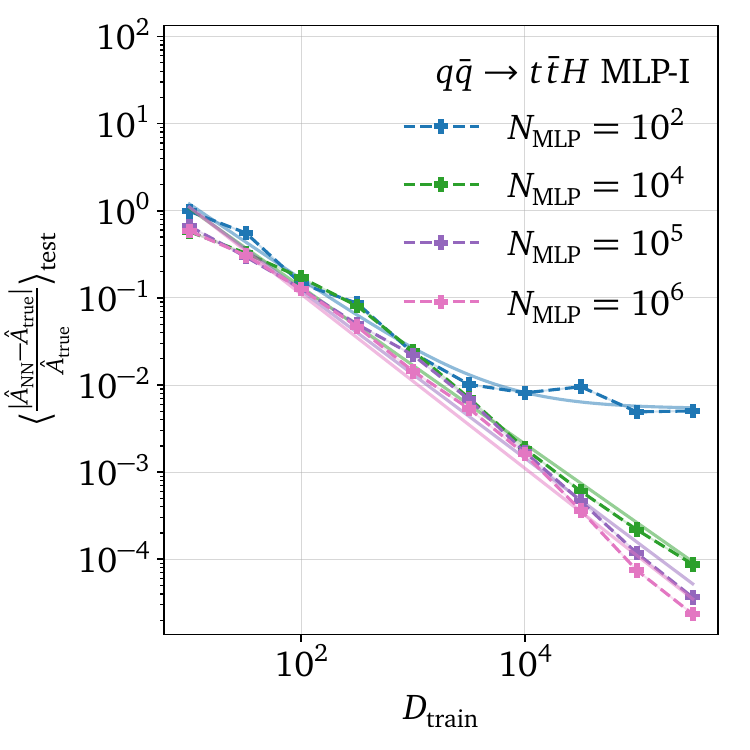}
    \includegraphics[width=0.49\linewidth]{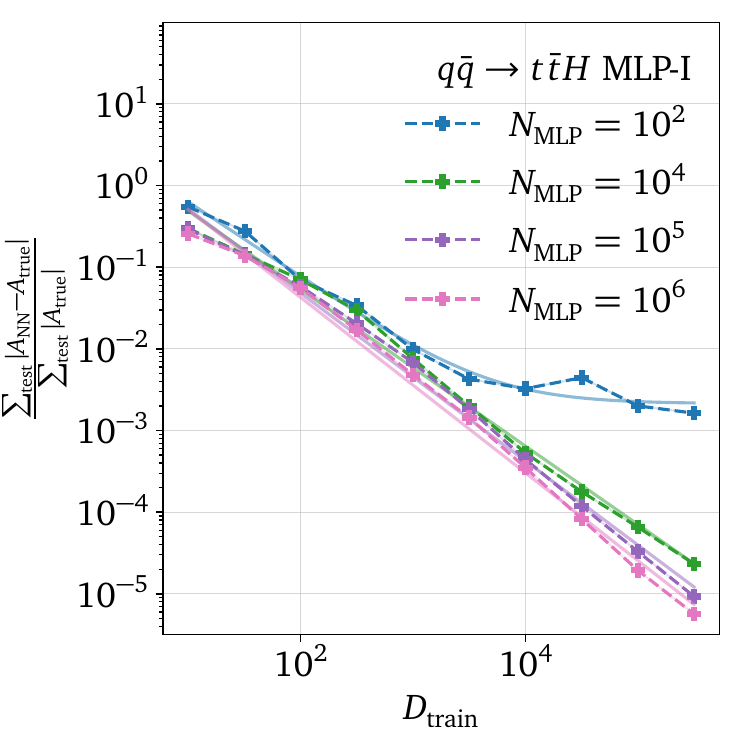}
    \caption{Left: Relative L1 error for the MLP-I model as a function of the training dataset size for different numbers of total parameters of the network. Right: Same as left but using the error metric studied in Ref.~\cite{Breso:2024jlt}.}
    \label{fig:qq_tth_MLP-I_L1_vs_MSE}
    \end{center}
\end{figure}

Above, we have shown all scaling laws depicting the mean-squared-error test loss on the $y$-axis. In Fig.~\ref{fig:qq_tth_MLP-I_L1_vs_MSE}, we demonstrate that analogous scaling laws also appear if looking at the L1 test error. In comparison to the left panel of Fig.~\ref{fig:qq_tth_mlp_dtrain_nsize}, the slopes of the curves are reduced roughly by a factor of two. This is in agreement with Eq.~\eqref{eq:aD_pred} given that the L1 loss scales with the error to the power of one. 

For comparison, we also show on the right panel of Fig.~\ref{fig:qq_tth_MLP-I_L1_vs_MSE} the scaling behavior for the $\epsilon$ metric used in Ref.~\cite{Breso:2024jlt}, which is defined by
\begin{align}
    \epsilon = \frac{\sum_i|A_\text{NN}(x_i) -A_\text{true}(x_i)|}{\sum_i A_\text{true}(x_i)} \;. 
\end{align}
Apart from the normalization, this corresponds to an L1 loss. Consequently, we observe the same slope of the loss curve as in the left panel. The slope of the curves is also in agreement with the results of Ref.~\cite{Breso:2024jlt} for $D_\text{train}
\lesssim 10^4$, above which they hit the saturation regime.

\subsection{Jet-associated \texorpdfstring{$Z$}{Z} production}
\label{app:Zjets}

\begin{figure}[H]
    \begin{center}
    \includegraphics[width=0.49\linewidth]{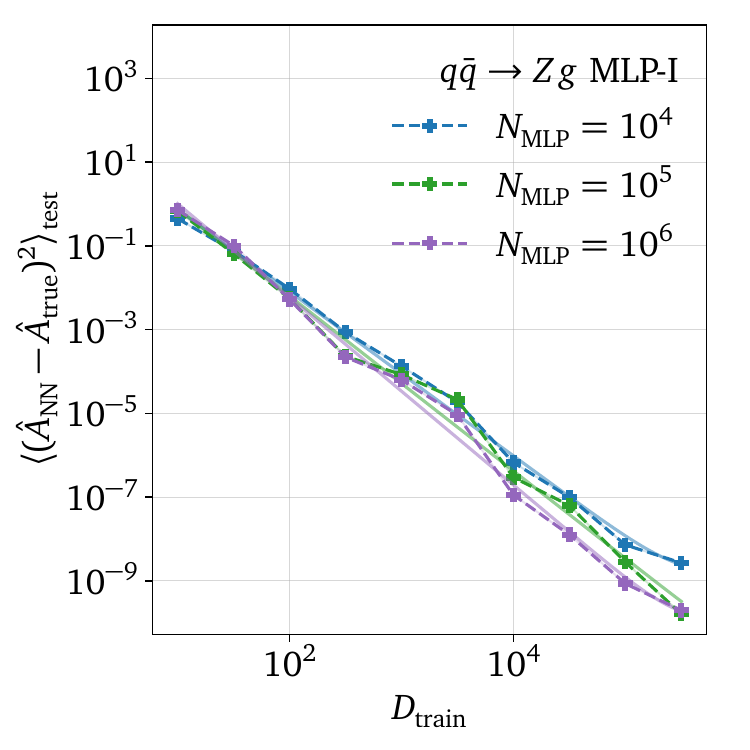}
    \includegraphics[width=0.49\linewidth]{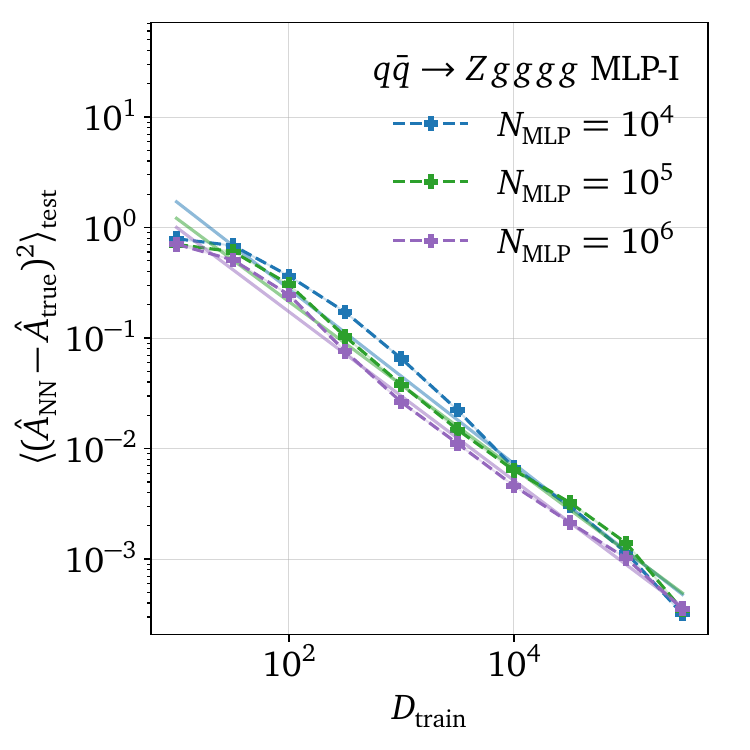}
    \caption{Left: MSE test errors for the $q\bar q\to Zg$ amplitude using the MLP-I model as a function of the size of the training dataset. Results are shown for three different numbers of trainable parameters. Right: Same as left, but for the $q\bar q \to Zgggg$ amplitude.}
    \label{fig:Zg_ND_NT}
    \end{center}
\end{figure}

In Fig.~\ref{fig:Zg_ND_NT}, we show the scaling of the $Zg$ and $Zgggg$ surrogates test losses as a function of the training dataset size for different NN sizes. We clearly see the emergence of scaling laws and confirm that the training dataset size is a major limiting factor in the considered parameter regime. The NN size only plays a minor role. We also observe that the $Zgggg$ is less precise as expected by the higher-dimensional phase space and consequently the slope of the loss curves is smaller than for the $Zg$ surrogates.

\begin{figure}[H]
    \begin{center}
    \includegraphics[width=0.49\linewidth]{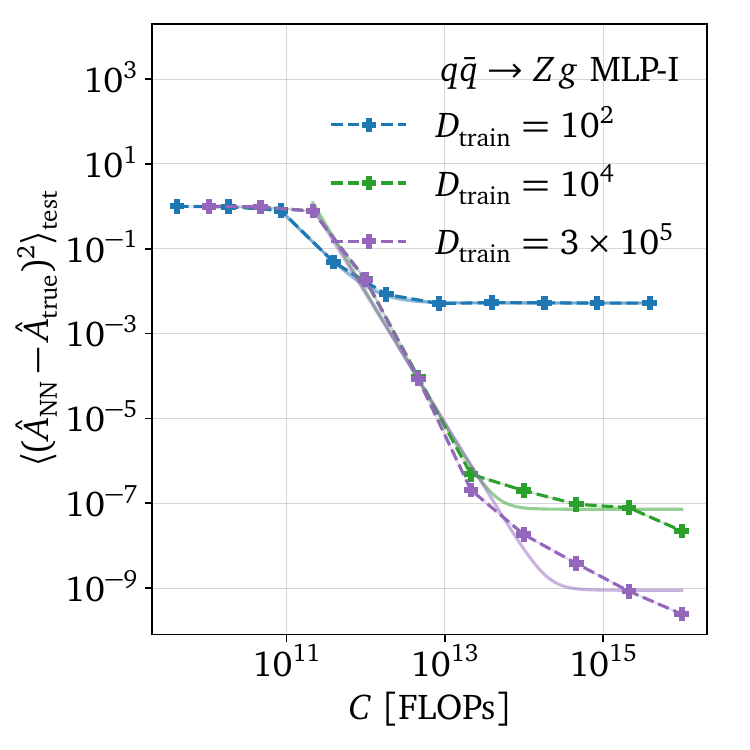}
    \includegraphics[width=0.49\linewidth]{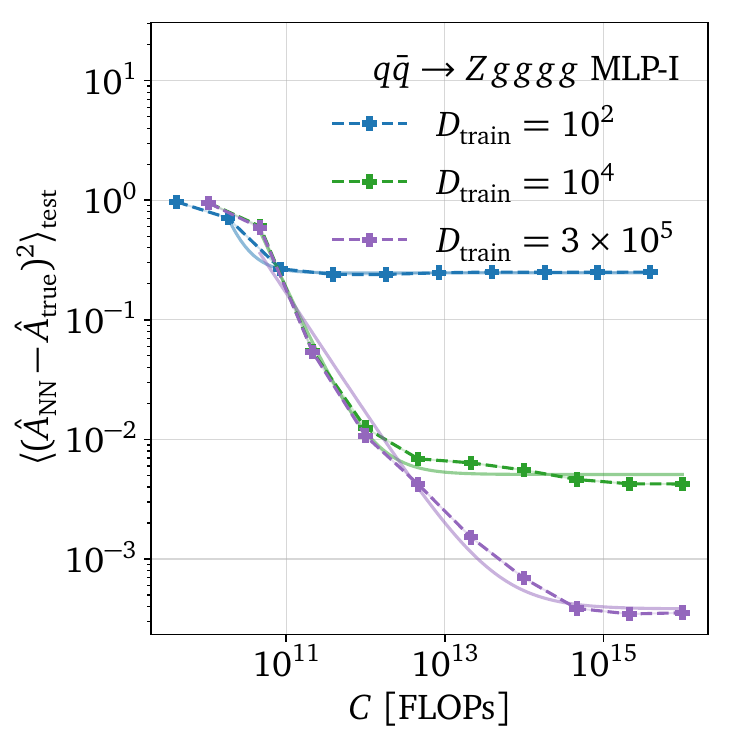}
     \caption{Left: MSE test errors for the $q\bar q\to Z g$ amplitude using the MLP-I model as a function of the compute spent for training in FLOPs for three different sizes of the training dataset. Right: Same as left, but for the $q\bar q \to Zgggg$ amplitude.}
    \label{fig:Zgggg_ND_NT}
    \end{center}
\end{figure}

The scaling as a function of the spent compute is shown in Fig.~\ref{fig:Zgggg_ND_NT} for different training dataset size. The training dataset size is again confirmed to be a major limiting factor.

\subsection{Jet associated di-photon production}
\label{app:diphoton}

\begin{figure}[H]
    \centering
    \includegraphics[width=0.49\linewidth]{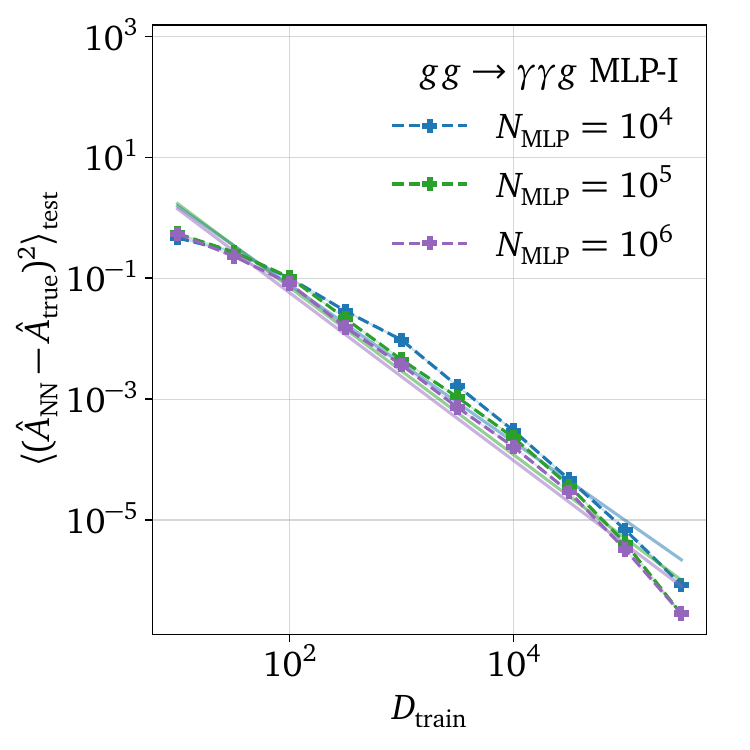}
    \includegraphics[width=0.49\linewidth]{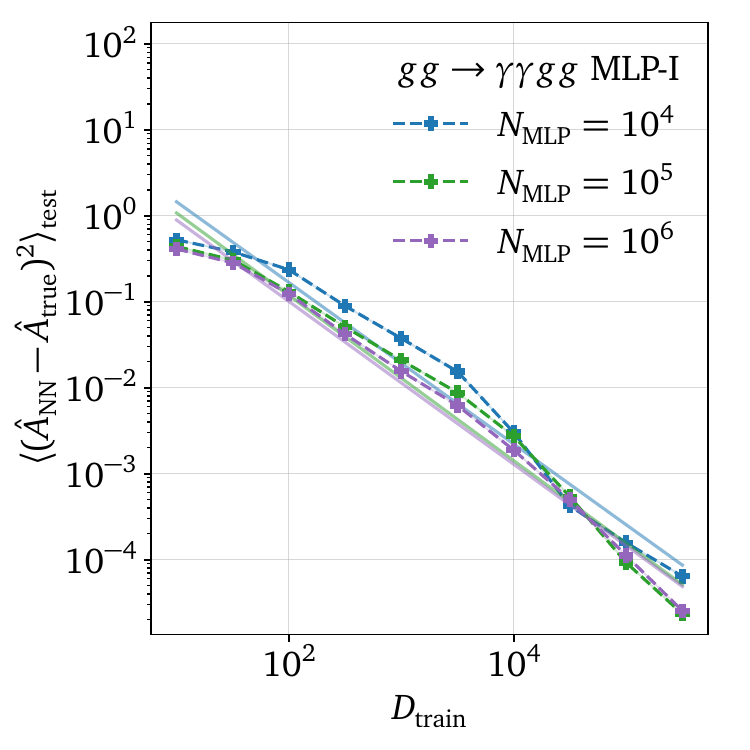}
    
    \caption{Left: MSE test errors for the $gg\to \gamma\gamma+g$ amplitude using the MLP-I model as a function of the size of the training dataset. Results are shown for three different numbers of trainable parameters. Right: Same as left, but for the $gg\to\gamma\gamma+gg$ amplitude.}
    \label{fig:aa_g_gg_ND}
\end{figure}

\begin{figure}[H]
    \centering
    \includegraphics[width=0.49\linewidth]{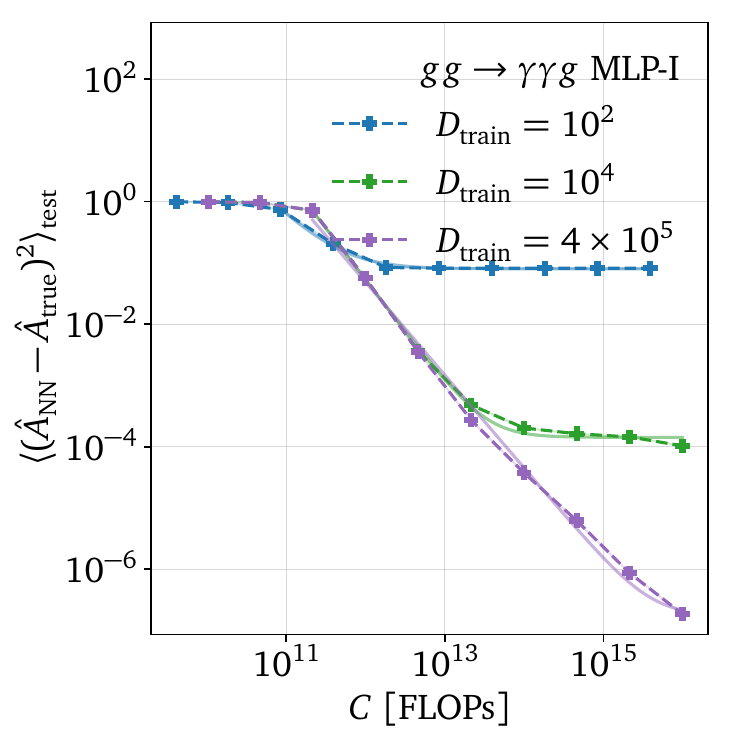}
    \includegraphics[width=0.49\linewidth]{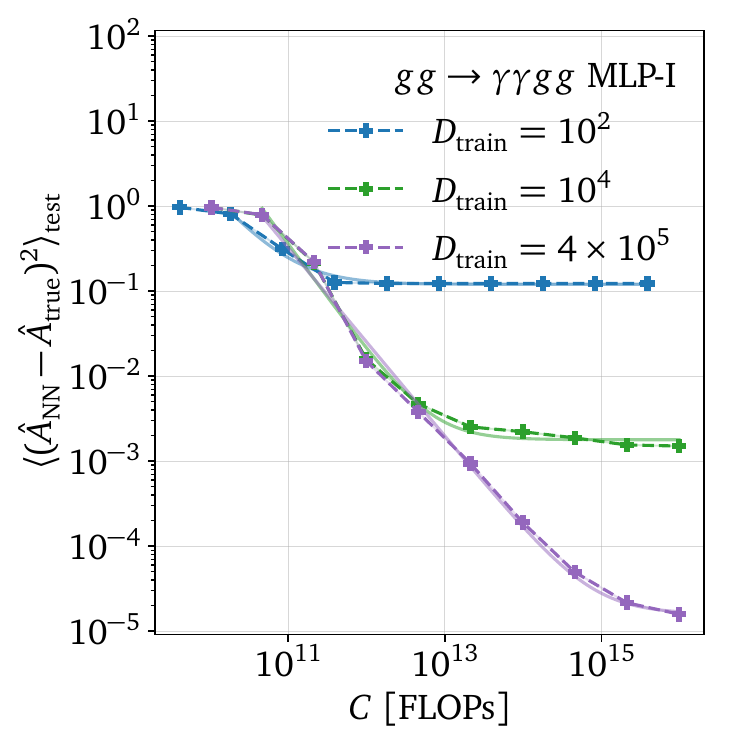}
    \caption{Left: MSE test errors for the $gg\to \gamma\gamma+g$ amplitude using the MLP-I model as a function of compute spent for training. Results are shown for three sizes of the training dataset. Right: Same as left, but for the $gg\to\gamma\gamma+gg$ amplitude.}
    \label{fig:aa_g_gg_NT}
\end{figure}

We show the scaling of the di-photon surrogates' test losses as function of the training dataset size in Fig.~\ref{fig:aa_g_gg_ND}. As for the other studied processes, the training dataset size is a major limiting factor with the NN size playing only a minor role. The scaling as a function of the spent compute shown in Fig.~\ref{fig:aa_g_gg_NT} for different training dataset sizes confirms this behavior.

\subsection{Electroweak multi-boson production}
\label{app:wz}

\begin{figure}[H]
    \begin{center}
    \includegraphics[width=0.49\linewidth]{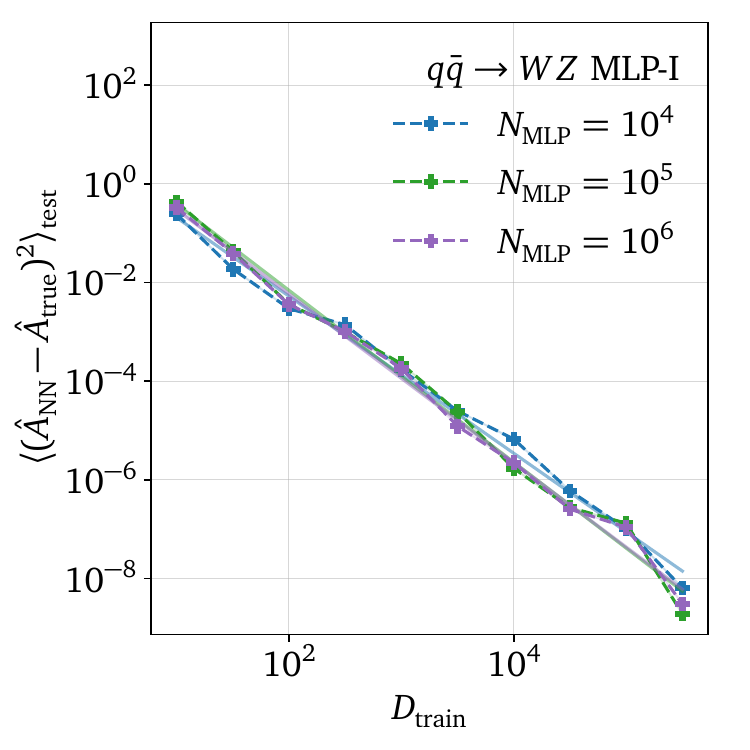}
    \includegraphics[width=0.49\linewidth]{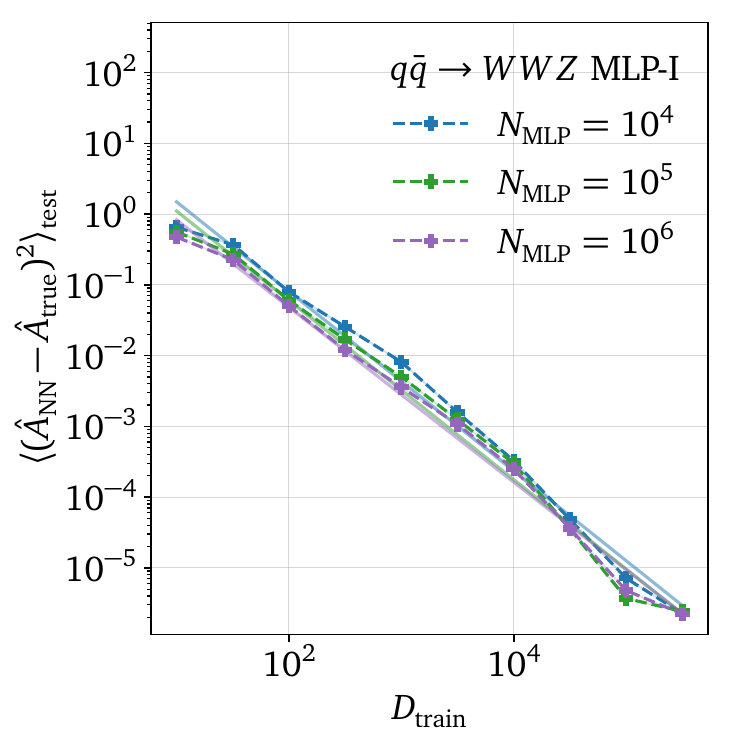}
    \includegraphics[width=0.49\linewidth]{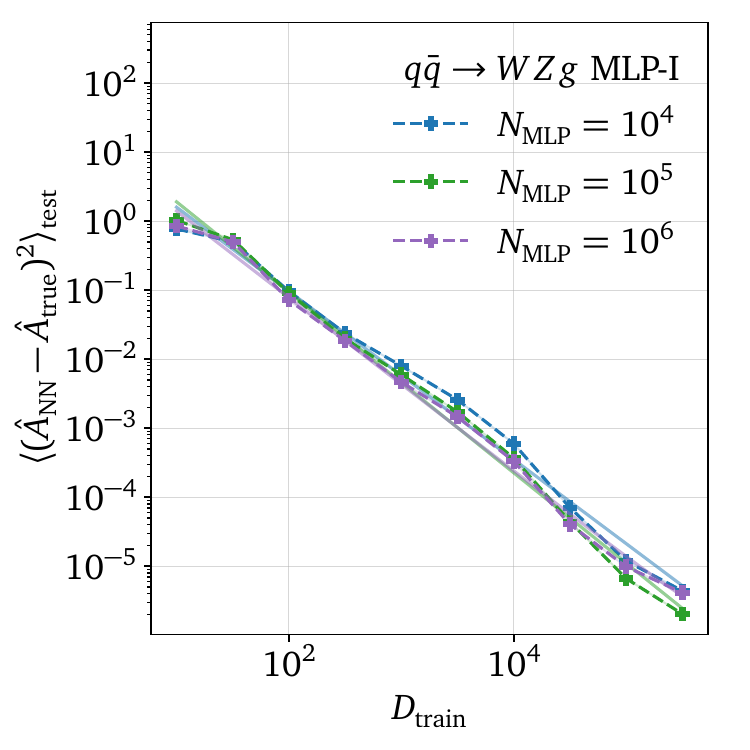}
    \includegraphics[width=0.49\linewidth]{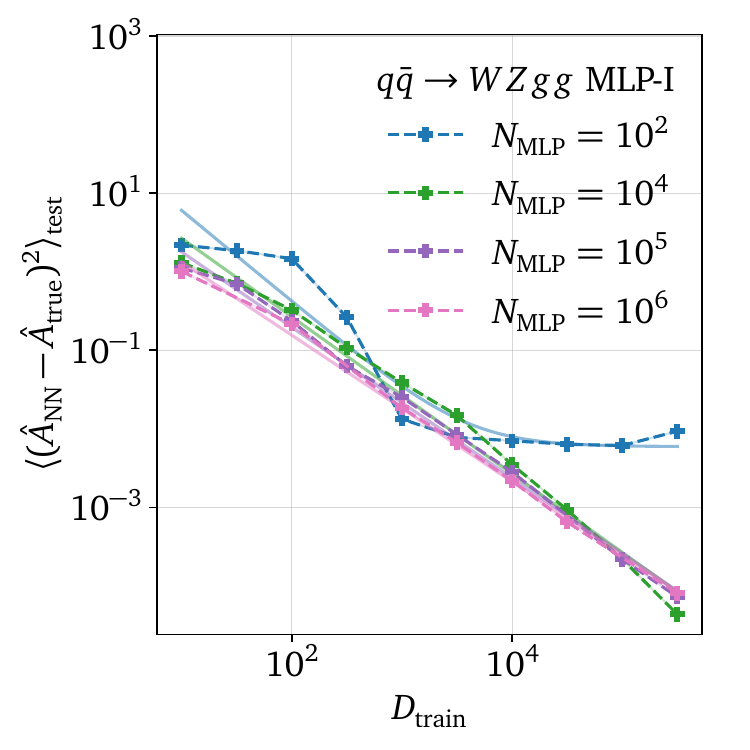}
    
    \caption{Upper left: MSE test errors for the $q\bar q\to WZ$ amplitude using the MLP-I model as a function of the size of the training dataset for different NN sizes. Results are shown for three different numbers of trainable parameters. Upper right: Same as upper left but for the $q\bar q \to WWZ$ amplitude. Lower left: Same as upper left but for the $q\bar q \to WZg$ amplitude. Lower right: Same as lower left but for the $q\bar q \to WZgg$ amplitude.}
    \label{fig:qq_wz_ND}
    \end{center}
\end{figure}

\begin{figure}[H]
    \begin{center}
    \includegraphics[width=0.49\linewidth]{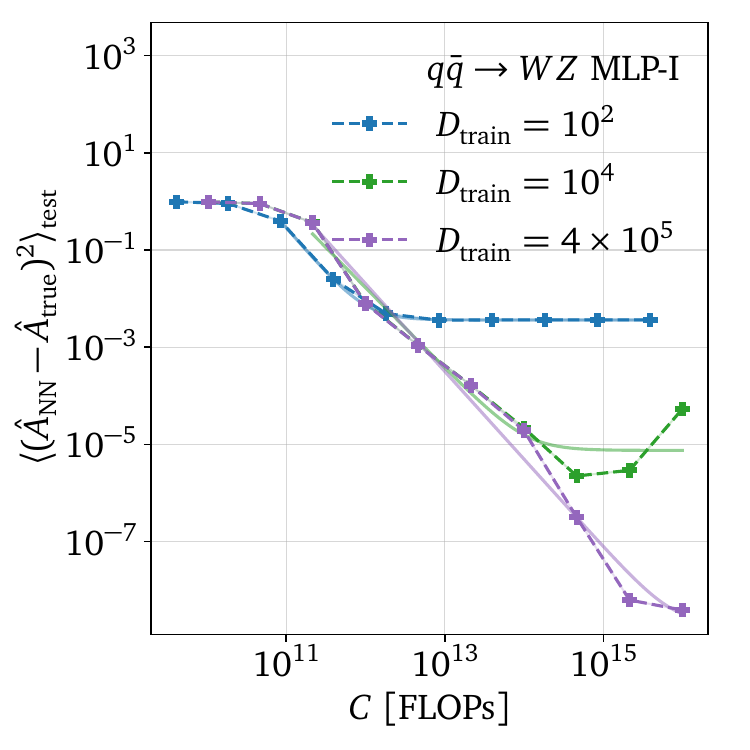}
    \includegraphics[width=0.49\linewidth]{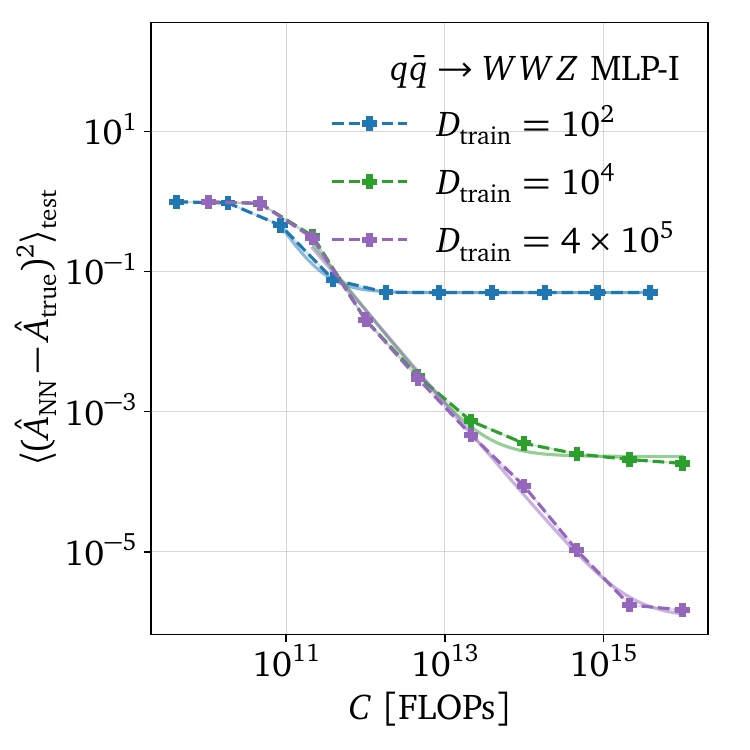}
    \includegraphics[width=0.49\linewidth]{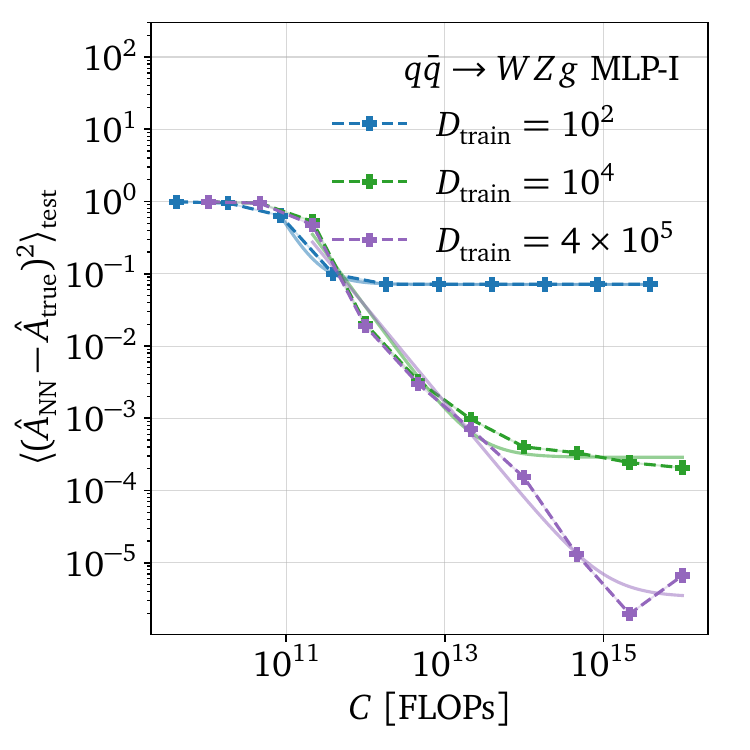}
    \includegraphics[width=0.49\linewidth]{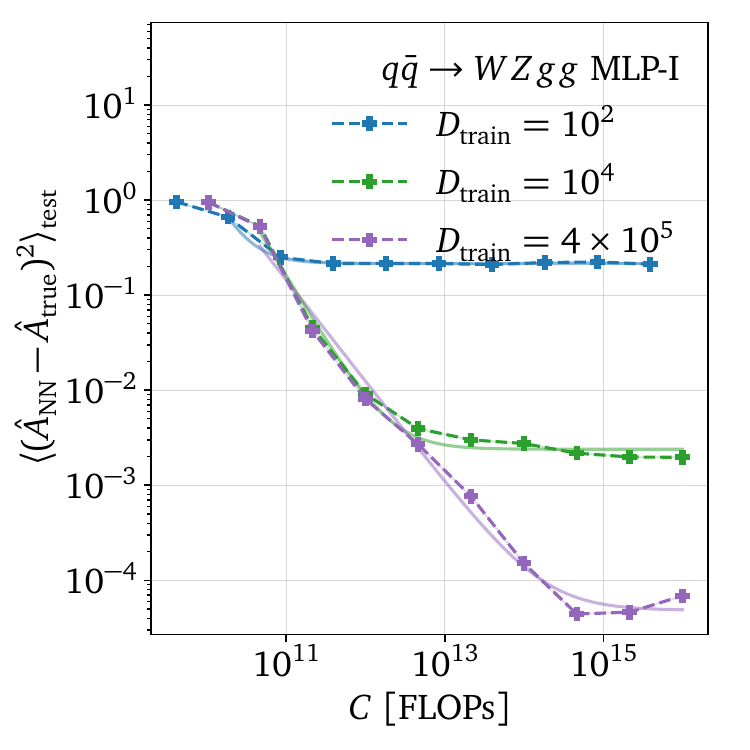}
    \caption{Upper left: MSE test errors for the $q\bar q\to WZ$ amplitude using the MLP-I model as a function of the compute spent for training for different sizes of the training dataset. Upper right: Same as left but for the $q\bar q \to WWZ$ amplitude. Lower left: Same as upper left but for the $q\bar q \to WZg$ amplitude. Lower right: Same as lower left but for the $q\bar q \to WZgg$ amplitude.}
    \label{fig:qq_wz_NT}
    \end{center}
\end{figure}

The MSE test errors of the electroweak multi-boson surrogates as a function of the training dataset size for different NN sizes are shown in Fig.~\ref{fig:qq_wz_ND}. As for the other processes, the training dataset size is an important limiting factor with the NN size playing only a minor role in the considered parameter regime. The scaling curves as a function of the compute for different training dataset sizes shown in Fig.~\ref{fig:qq_wz_NT} confirm this behavior. For some of the curves --- e.g., the $D_\text{train} = 4\times 10^5$ curve in the lower panel of Fig.~\ref{fig:qq_wz_NT} ---, the test loss slightly increases when going from the second-to-last to the last datapoint when raising $C$. We found this behavior to be due to a not-fine enough scan search for the optimal learning rate. While the learning rate is optimized for the highest compute point, the not-fine enough optimization resulted in the learning rate actually being optimal for the second-to-last point creating the observed behavior. A finer learning-rate scan at this point would remove this feature. However, a sufficiently fine, independent learning rate scan for every $(\text{process}, D_\text{train}, C)$ combination lies beyond our compute budget.

\section{Intrinsic dimension}
\label{app:intrinsic_dimension}

\subsection{Method}

For estimating the intrinsic dimension of the representation learned by the network, we use the twoNN method outlined in Refs.~\cite{NIPS2004_74934548,2017NatSR...712140F,2019arXiv190512784A}. 

Given a point $x_i$ in a dataset, we build a list of the $k$ nearest neighbours with $r_1$, $r_2$, $\ldots$, $r_k$ being the sorted distances. Then, the volume of the hyperspherical shell between two successive neighbors $l-1$ and $l$ is given by
\begin{align}
    v_l = \omega_d (r_l^d - r_{l-1}^d)\;,
\end{align}
where $\omega_d$ is the volume of the $d$-dimension unit-sphere. If the point density $\rho$ around $x_i$ is constant, the volumes $v_l$ are distributed according to an exponential distribution with the rate being equal to $\rho$,
\begin{align}
    p(v_l\in[v, v+dv]) = \rho e^{-\rho v}dv\;.
\end{align}
This knowledge allows us to obtain the distribution of the ratio
\begin{align}
    R_{ij}  \equiv \frac{v_i}{v_j}
\end{align}
between two shell volumes as
\begin{align}
    p(R_{ij}\in[R, R + dR]) &= \int_0^\infty dv_i\int_0^\infty dv_j\, \rho^2 e^{-\rho(v_i+v_j)}\mathbf{1}_{\left\{\frac{v_i}{v_j}\in [R, R+dR]\right\}} = \notag\\
    &= \frac{\rho^2 dR}{(1+R)^2}\cdot\int_0^\infty du\, u\, e^{-\rho u} = \notag\\
    &= \frac{dR}{(1 + R)^2}\;, \label{eq:pR_dist}
\end{align}
where we have used the variable $u= v_i + v_j$ and $\mathbf{1}$ is the indicator function. Defining
\begin{align}
    \mu \equiv \frac{r_2}{r_1}\;,
\end{align}
we can rewrite $R_{21}$ as
\begin{align}
    R_{21} = \frac{v_2}{v_1} = \frac{r_2^d - r_1^d}{r_1^d - r_0^d} = \mu^d - 1\;.
\end{align}
Using Eq.~\eqref{eq:pR_dist}, the distribution of $\mu$ is given by
\begin{align}
    p(\mu|d) = d\,\mu^{-d -1}\;,
\end{align}
which has the form of a Pareto distribution. From this expression, we can estimate the value of $d$ as the one that maximizes the joint likelihood 
\begin{equation}
    P(\mathbf{\mu}|d) = d^N \prod_{i=1}^N \mu_{i}^{-(d+1)},
\end{equation}
where $N$ is the total number of points in the dataset and $\mathbf{\mu}=(\mu_1,\ldots,\mu_N)$. 

To approximate the intrinsic dimension used by the NN, we consider the activations of the last hidden layers $\{x_i\}$ accumulated over a subset of the full training dataset with a size of at least $10^4$ phase-space points. $\mu$ is then approximated by computing for each $x_i$ in our dataset the distance to the nearest and second-nearest neighbors. Finally, we maximize $P(\mathbf{\mu}|d)$ by using the code implementation attached to Ref.~\cite{ansuini2019intrinsic}, which leverages the empirical cumulate of the distribution to reduce the estimation of $d$ to a linear regression problem. 

\subsection{Results}

\begin{figure}[H]
    \begin{center}
    \includegraphics[width=0.49\linewidth]{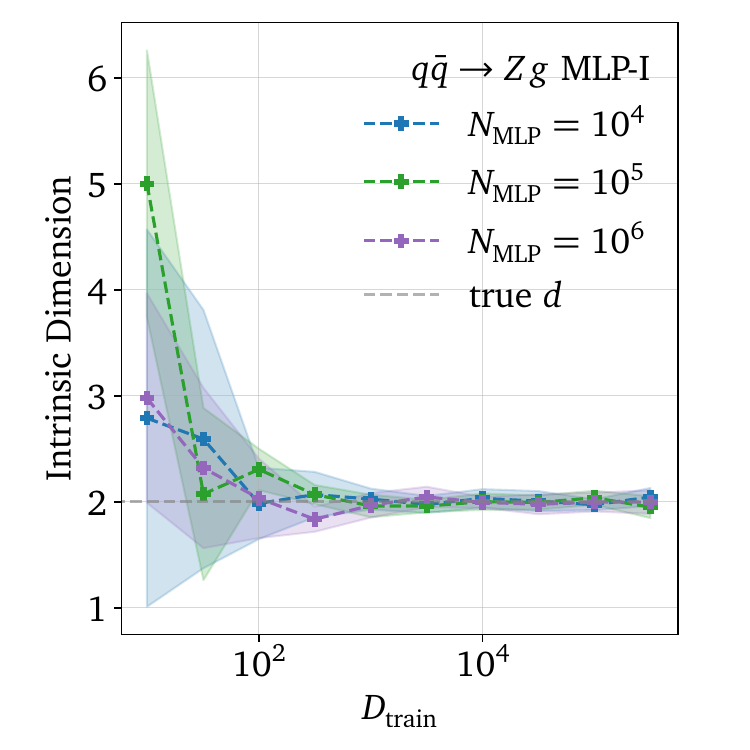}
    \includegraphics[width=0.49\linewidth]{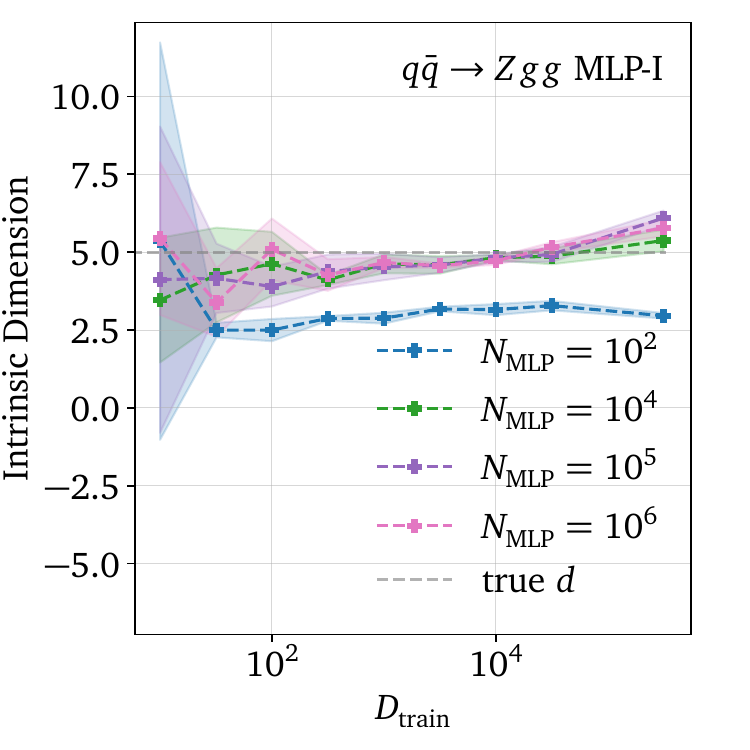}
    \includegraphics[width=0.49\linewidth]{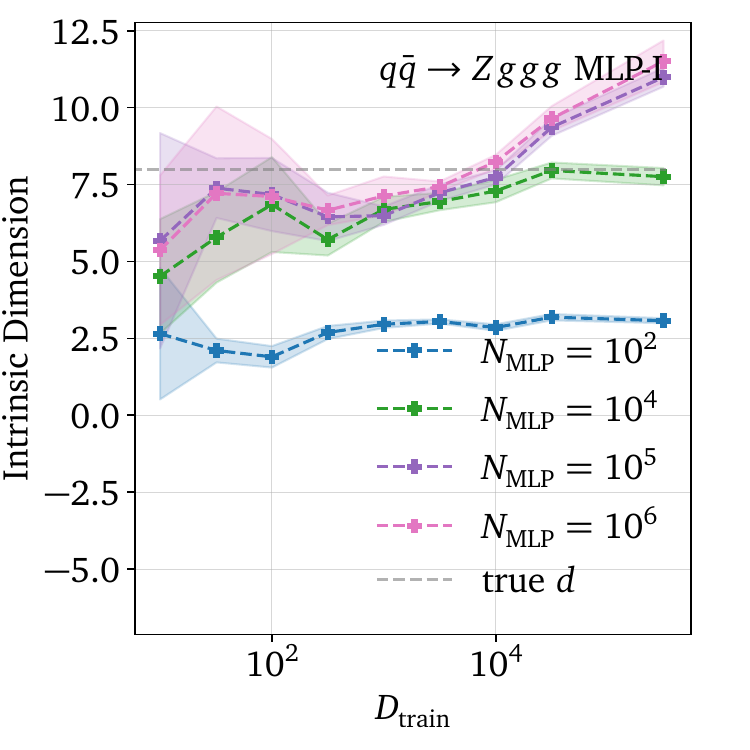}
    \includegraphics[width=0.49\linewidth]{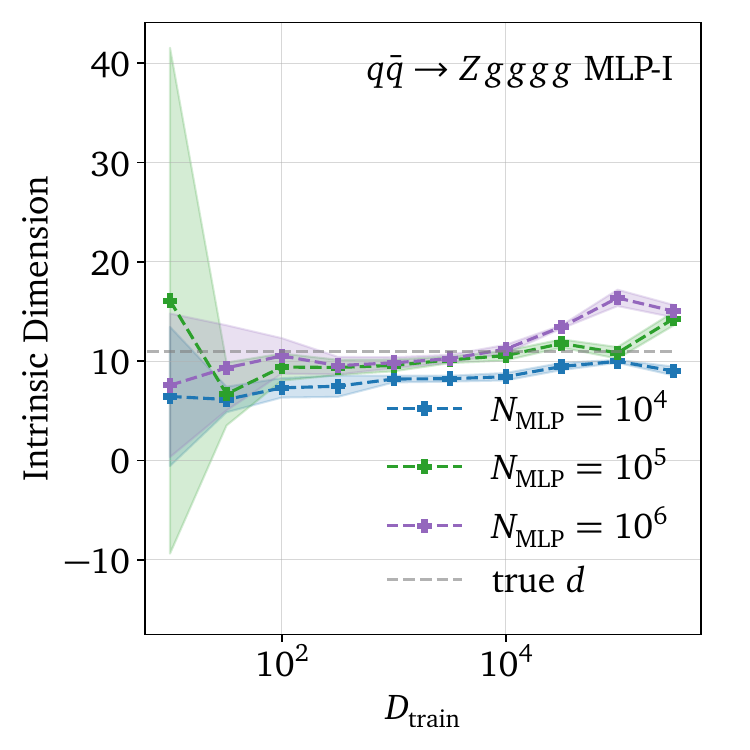}
    \caption{Intrinsic dimension extracted by the $q\bar q\to Z+\text{gluons}$ surrogates as a function of the training dataset size for different NN sizes. The color bounds indicate the associated uncertainty. The intrinsic dimension is calculated based on a subset of the final-layer activations with the uncertainties corresponding to the standard deviation obtained by sampling different subsets.}
    \label{fig:intr_dim_z}
    \end{center}
\end{figure}

We show the intrinsic dimension used by the $Z + \text{gluons}$ surrogates in Fig.~\ref{fig:intr_dim_z}. We observe that the true dimensions of the phase spaces are reasonably well extracted if the NN and training dataset size are large enough. Only for the $Zggg$ and $Zgggg$ surrogates, the deviations from the expected values are sizeable but overall still small.

\begin{figure}[H]
    \begin{center}
    \includegraphics[width=0.49\linewidth]{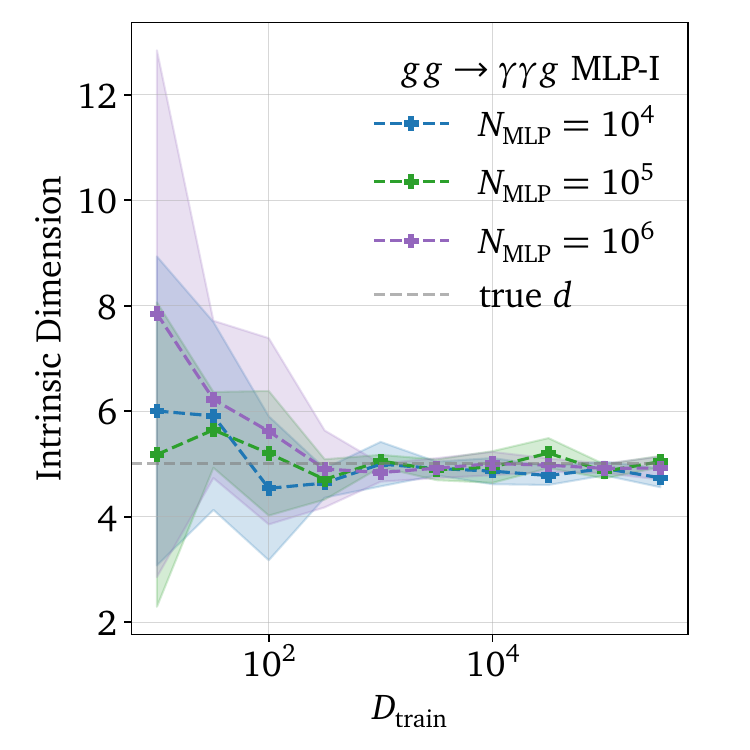}
    \includegraphics[width=0.49\linewidth]{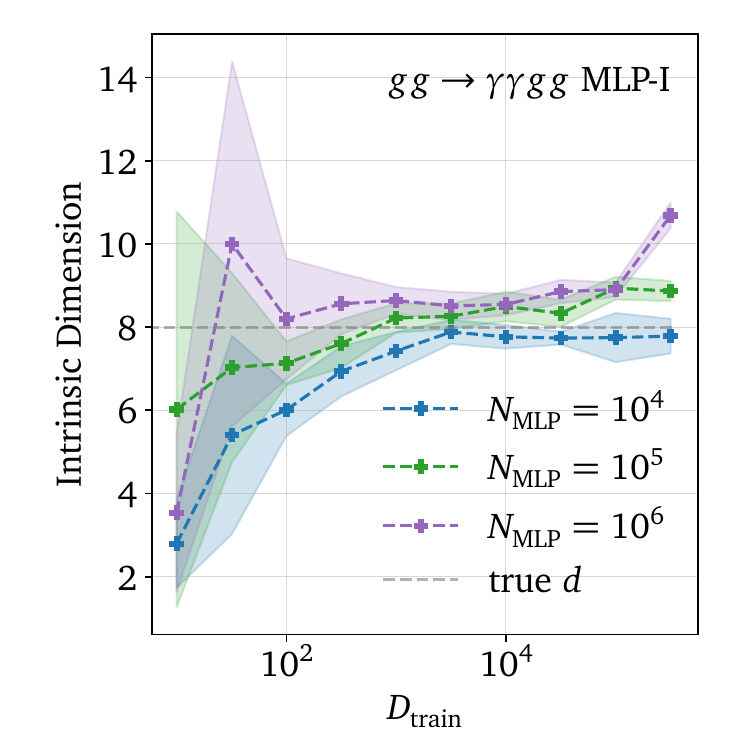}
    \caption{Intrinsic dimension extracted by the $gg\to \gamma\gamma g, \gamma\gamma gg$ surrogates as a function of the training dataset size for different NN sizes. The color bounds indicate the associated uncertainty. The intrinsic dimension is calculated based on a subset of the final-layer activations with the uncertainties corresponding to the standard deviation obtained by sampling different subsets.}
    \label{fig:intr_dim_aa}
    \end{center}
\end{figure}

We find a similar behaviour for the di-photon surrogates shown in Fig.~\ref{fig:intr_dim_aa}. The overall agreement with the theoretical expectation is quite good. For the higher-dimensional $gg\to \gamma\gamma gg$ process, the surrogates struggle a bit to reduce the phase space to its underlying minimal dimension.

\begin{figure}[H]
    \begin{center}
    \includegraphics[width=0.49\linewidth]{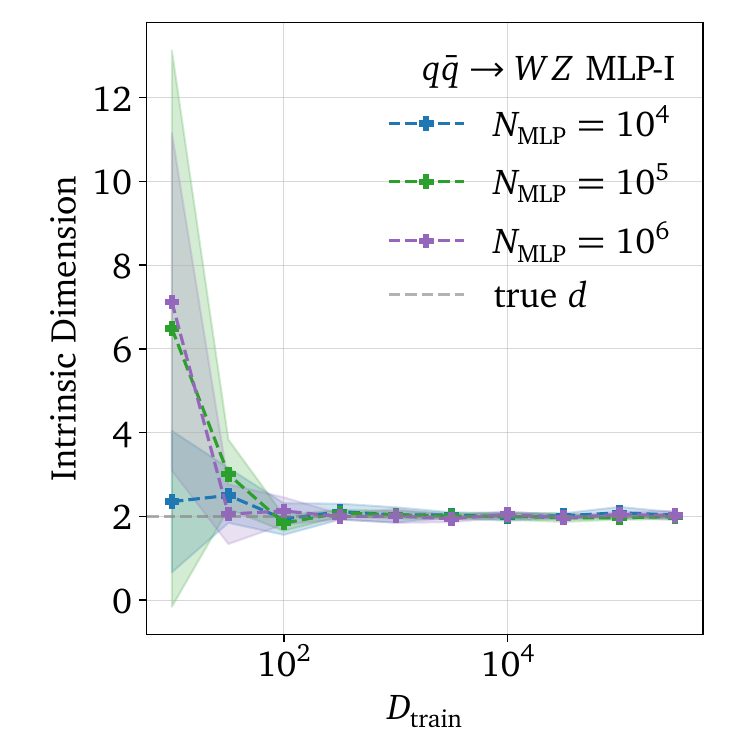}
    \includegraphics[width=0.49\linewidth]{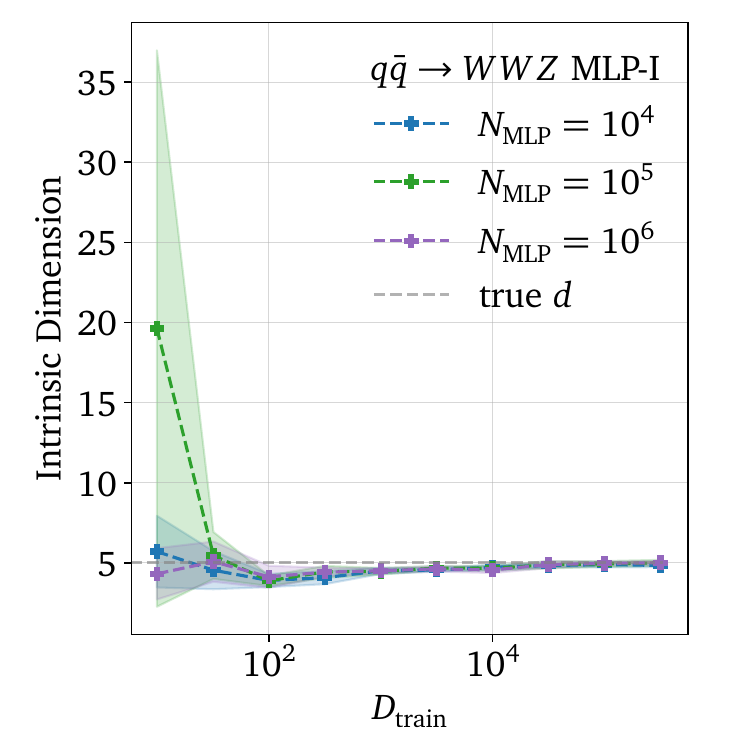}
    \includegraphics[width=0.49\linewidth]{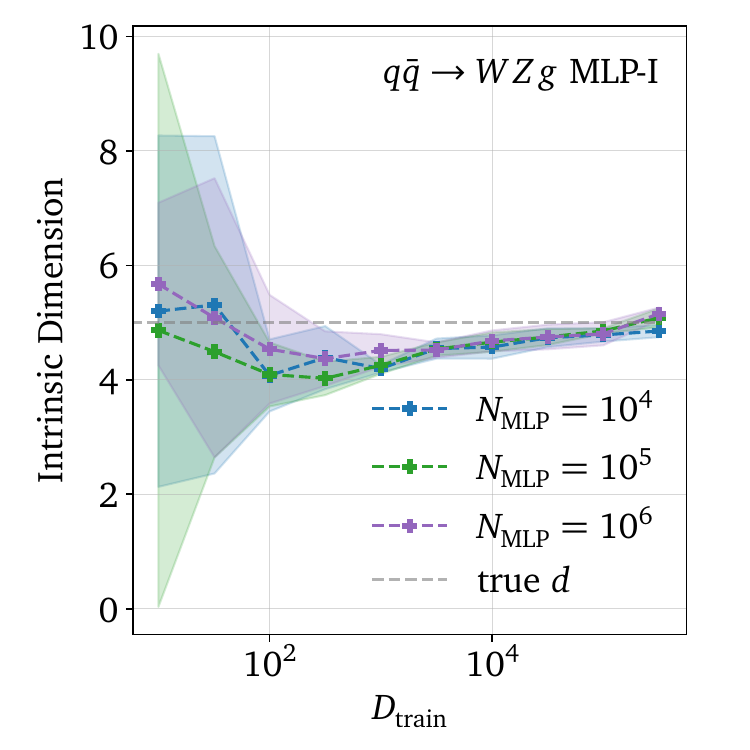}
    \caption{Intrinsic dimension extracted by the $q\bar q\to WZ, WWZ,WZg$ surrogates as a function of the training dataset size for different NN sizes. The color bounds indicate the associated uncertainty. The intrinsic dimension is calculated based on a subset of the final-layer activations with the uncertainties corresponding to the standard deviation obtained by sampling different subsets.}
    \label{fig:intr_dim_wz}
    \end{center}
\end{figure}

For the electroweak multi-boson processes shown in Fig.~\ref{fig:intr_dim_wz}, we again find an excellent convergence towards the theoretically expected values. 

All of these results hold even if we train the MLP-I with 4-momenta inputs instead of the momentum invariants, signaling that the intrinsic dimension can be estimated even if we are not fully aware of the symmetries of the process. Thus, it is clear that the twoNN method applied on trained MLP-I surrogates can be used reliably to recover the number of degrees of freedom of a wide range of processes.

\section{Power law fit results}
\label{app:fits}

In this Appendix together with Tab.~\ref{tab:paramfit_MLP}, we provide the fitted power-law coefficients for all processes and models studied in the paper. The uncertainties are derived from the fitting procedure assuming a 10\% relative uncertainty on the test loss values. This is a conservative estimate based on the results of Ref.~\cite{Breso:2024jlt}. The uncertainties are supposed to capture the variation in the fitted coefficients from rerunning the same NN with the same hyperparameter settings. They do not cover the potentially achievable improvements from further optimizing the hyperparameters. As we did for Tab.~\ref{tab:paramfit_MLP}, we approximate $K_X$ to 0 whenever the uncertainty from the fit appears unbounded from below.

\begin{table}[H]
    \centering
    \renewcommand{\arraystretch}{1.3}
    \begin{tabular}{l@{\hskip 15pt} c@{\hskip 15pt} c@{\hskip 15pt} c@{\hskip 15pt} c}
        \toprule
        param.\ $X$ &  2nd param. & $X_c$ & $\alpha_X$ & $K_X$ \\
        \midrule
$D_\text{train}$ & $N_\text{MLP}$ = $10^{4}$ & $\left(6.98 ^{+0.56 }_{-0.31}\right)\times 10^{1}$ & $1.752 \pm 0.028$ & $\left(2.23 ^{+387}_{-0.22}\right)\times 10^{-7}$ \\
$D_\text{train}$ & $N_\text{MLP}$ = $10^{5}$ & $\left(2.55^{+0.23}_{-0.12}\right)\times 10^{1}$ & $1.949 \pm 0.030$ & $\left(4.86 ^{+7.3}_{-0.46}\right)\times 10^{-8}$ \\
$D_\text{train}$ & $N_\text{MLP}$ = $10^{6}$ & $\left(2.073 ^{+0.18}_{-0.096 }\right)\times 10^{1}$ & $2.239 \pm 0.051$ & $\left(3.10 ^{+0.76}_{-0.22}\right) \times 10^{-6}$ \\
\hline
$C$ [FLOPs] & $D_\text{train}$ = $10^{2}$ & $\left(6.11  ^{+0.80 }_{-0.35}\right)\times 10^{10}$ & $1.89 \pm 0.94$ & $\left(3.11 ^{+0.48}_{-0.19}\right)\times 10^{-2}$ \\
$C$ [FLOPs] & $D_\text{train}$ = $10^{4}$ & $\left(12.43  ^{+0.81}_{-0.49}\right)\times 10^{10}$ & $2.13 \pm 0.037$ & $\left(2.21 ^{+0.54}_{-0.16}\right)\times 10^{-6}$ \\
$C$ [FLOPs] & $D_\text{train}$ = $10^{6}$ & $\left(11.95  ^{+0.81}_{-0.48}\right)\times 10^{10}$ & $2.017 \pm 0.027$ & $\left(4.92 ^{+1.9}_{-0.39}\right)\times 10^{-8}$ \\       
        \bottomrule
    \end{tabular}
    \caption{Fitted scaling parameters for $t\bar{t}H$ MLP-I surrogates trained with heteroscedastic loss.}
    \label{tab:paramfit_qq_tth_MLP_HETEROSC}
\end{table}

\begin{table}[H]
    \centering
    \renewcommand{\arraystretch}{1.3}
    \begin{tabular}{l@{\hskip 15pt} c@{\hskip 15pt} c@{\hskip 15pt} c@{\hskip 15pt} c}
        \toprule
        param.\ $X$ &  2nd param. & $X_c$ & $\alpha_X$ & $K_X$ \\
        \midrule
        
$D_\text{train}$ & $N_\text{LLoCa-Tr}$ = $6\times10^{6}$&  $6.09^{+0.54}_{-0.29}$ & $1.303 \pm 0.015 $ & $\simeq 0$ \\                  
        \bottomrule
    \end{tabular}
    \caption{Fitted scaling parameters for $t\bar{t}H$ LLoCa-Transformer surrogates trained with MSE loss.}
    \label{tab:paramfit_qq_tth_lloca_MSE}
\end{table}

\begin{table}[H]
    \centering
    \renewcommand{\arraystretch}{1.3}
    \begin{tabular}{l@{\hskip 15pt} c@{\hskip 15pt} c@{\hskip 15pt} c@{\hskip 15pt} c}
        \toprule
        param.\ $X$ &  2nd param. & $X_c$ & $\alpha_X$ & $K_X$ \\
        \midrule
\hline
\multirow{3}{*}{$D_\text{train}$}
 & $N_\text{MLP}$ = $10^{4}$ & $8.55 ^{+0.96}_{-0.45}$ & $1.958 \pm 0.028$ & $\left(1.31^{+14}_{-0.13}\right)\times 10^{-9}$ \\
 & $N_\text{MLP}$ = $10^{5}$ & $8.73 ^{+0.89}_{-0.44}$ & $2.084 \pm 0.027$ & $\simeq 0$ \\
 & $N_\text{MLP}$ = $10^{6}$ & $10.1 ^{+0.91}_{-0.48}$ & $2.243 \pm 0.027$ & $\left(1.61 ^{+2.4}_{-0.15 }\right)\times 10^{-10}$ \\
 \hline
 \multirow{3}{*}{$C$ [FLOPs]}
 & $D_\text{train}$ = $10^{2}$ & $\left(7.38^{+2.0}_{-0.54}\right)\times 10^{10}$ & $1.84 \pm 0.18$ & $\left(5.09 ^{+0.91}_{-0.33}\right)\times 10^{-3}$ \\
 & $D_\text{train}$ = $10^{4}$ & $\left(2.28 ^{+0.19 }_{-0.10}\right)\times 10^{11}$ & $3.129 \pm 0.070$ & $\left(7.21 ^{+1.6}_{-0.50 }\right)\times 10^{-8}$ \\
 & $D_\text{train}$ = $3.03 \times 10^{5}$ & $\left(20.97 ^{+1.7}_{-0.94 }\right)\times 10^{10}$ & $3.039 \pm 0.050$ & $\left(8.88 ^{+2.6 }_{-0.66}\right) \times 10^{-10}$ \\
 \bottomrule
    \end{tabular}
    \caption{Fitted scaling parameters for $Zg$ MLP-I surrogates trained with MSE loss.}
    \label{tab:paramfit_zg_MLP_MSE}
\end{table}

\begin{table}[H]
    \centering
    \renewcommand{\arraystretch}{1.3}
    \begin{tabular}{l@{\hskip 15pt} c@{\hskip 15pt} c@{\hskip 15pt} c@{\hskip 15pt} c}
        \toprule
        param.\ $X$ &  2nd param. & $X_c$ & $\alpha_X$ & $K_X$ \\
        \midrule
\hline
\multirow{4}{*}{$D_\text{train}$}
 & $N_\text{MLP}$ = $10^{2}$ & $\left(2.65^{+0.43}_{-0.16}\right)\times 10^{1}$ & $1.373 \pm 0.055$ & $\left(6.81^{+1.9}_{-0.50}\right)\times 10^{-4}$ \\
 & $N_\text{MLP}$ = $10^{4}$ & $15.21^{+2.7}_{-0.97}$ & $1.316 \pm 0.029$ & $\simeq 0$ \\
 & $N_\text{MLP}$ = $10^{5}$ & $10.92^{+2.3}_{-0.74}$ & $1.301 \pm 0.031$ & $\left(3.24^{+11}_{-0.32}\right)\times 10^{-6}$ \\
 & $N_\text{MLP}$ = $10^{6}$ & $8.68^{+2.0}_{-0.60}$ & $1.280 \pm 0.031$ & $\left(2.50^{+12}_{-0.25}\right)\times 10^{-6}$ \\\hline
\multirow{3}{*}{$C$ [FLOPs]}
 & $D_\text{train}$ = $10^{2}$ & $\left(1.65^{+1.1}_{-0.14}\right)\times 10^{10}$ & $1.25 \pm 0.21$ & $\left(4.70^{+0.76}_{-0.29}\right)\times 10^{-2}$ \\
 & $D_\text{train}$ = $10^{4}$ & $\left(3.60^{+0.98}_{-0.26}\right)\times 10^{10}$ & $1.600 \pm 0.074$ & $\left(12.66^{+2.6}_{-0.85}\right)\times 10^{-5}$ \\
 & $D_\text{train}$ = $10^{5}$ & $\left(3.04^{+0.74}_{-0.22}\right)\times 10^{10}$ & $1.476 \pm 0.039$ & $\left(1.70^{+0.60}_{-0.13}\right)\times 10^{-6}$ \\                   
        \bottomrule
    \end{tabular}
    \caption{Fitted scaling parameters for $Zgg$ MLP-I surrogates trained with MSE loss.}
    \label{tab:paramfit_zgg_MLP_MSE}
\end{table}

\begin{table}[H]
    \centering
    \renewcommand{\arraystretch}{1.3}
    \begin{tabular}{l@{\hskip 15pt} c@{\hskip 15pt} c@{\hskip 15pt} c@{\hskip 15pt} c}
        \toprule
        param.\ $X$ &  2nd param. & $X_c$ & $\alpha_X$ & $K_X$ \\
        \midrule
\hline
\multirow{4}{*}{$D_\text{train}$}
 & $N_\text{MLP}$ = $10^{2}$ & $\left(5.00^{+0.81}_{-0.31}\right)\times 10^{1}$ & $1.193 \pm 0.062$ & $\left(6.68^{+1.8}_{-0.48}\right)\times 10^{-3}$ \\
 & $N_\text{MLP}$ = $10^{4}$ & $\left(2.32^{+0.57}_{-0.17}\right)\times 10^{1}$ & $1.030 \pm 0.030$ & $\simeq 0$ \\
 & $N_\text{MLP}$ = $10^{5}$ & $\left(1.59^{+0.46}_{-0.12}\right)\times 10^{1}$ & $1.008 \pm 0.030$ & $\simeq 0$ \\
 & $N_\text{MLP}$ = $10^{6}$ & $\left(1.268^{+0.40}_{-0.096}\right)\times 10^{1}$ & $0.997 \pm 0.031$ & $\simeq 0$ \\\hline
\multirow{3}{*}{$C$ [FLOPs]}
 & $D_\text{train}$ = $10^{2}$ & $\left(1.34^{+1.0}_{-0.12}\right)\times 10^{10}$ & $1.70 \pm 0.56$ & $\left(10.58^{+1.5}_{-0.63}\right)\times 10^{-2}$ \\
 & $D_\text{train}$ = $10^{4}$ & $\left(2.93^{+1.1}_{-0.23}\right)\times 10^{10}$ & $1.65 \pm 0.13$ & $\left(1.68^{+0.28}_{-0.11}\right)\times 10^{-3}$ \\
 & $D_\text{train}$ = $10^{5}$ & $\left(1.38^{+0.84}_{-0.12}\right)\times 10^{10}$ & $1.099 \pm 0.042$ & $\left(2.96^{+1.1}_{-0.23}\right)\times 10^{-5}$ \\
        \bottomrule
    \end{tabular}
    \caption{Fitted scaling parameters for $Zggg$ MLP-I surrogates trained with MSE loss.}
    \label{tab:paramfit_zggg_MLP_MSE}
\end{table}

\begin{table}[H]
    \centering
    \renewcommand{\arraystretch}{1.3}
    \begin{tabular}{l@{\hskip 15pt} c@{\hskip 15pt} c@{\hskip 15pt} c@{\hskip 15pt} c}
        \toprule
        param.\ $X$ &  2nd param. & $X_c$ & $\alpha_X$ & $K_X$ \\
        \midrule
\hline
\multirow{3}{*}{$D_\text{train}$}
 & $N_\text{MLP}$ = $10^{4}$ & $\left(1.98^{+0.85}_{-0.16}\right)\times 10^{1}$ & $0.788 \pm 0.031$ & $\simeq 0$ \\
 & $N_\text{MLP}$ = $10^{5}$ & $\left(1.39^{+0.72}_{-0.12}\right)\times 10^{1}$ & $0.774 \pm 0.032$ & $\simeq 0$ \\
 & $N_\text{MLP}$ = $10^{6}$ & $\left(1.010^{+0.62}_{-0.087}\right)\times 10^{1}$ & $0.765 \pm 0.034$ & $\simeq 0$ \\\hline
\multirow{3}{*}{$C$ [FLOPs]}
 & $D_\text{train}$ = $10^{2}$ & $\left(1.23^{+7.0}_{-0.12}\right)\times 10^{10}$ & $1.86 \pm 0.16$ & $\left(2.39^{+0.34}_{-0.14}\right)\times 10^{-1}$ \\
 & $D_\text{train}$ = $10^{4}$ & $\left(3.07^{+1.6}_{-0.26}\right)\times 10^{10}$ & $1.45 \pm 0.14$ & $\left(5.09^{+0.85}_{-0.32}\right)\times 10^{-3}$ \\
 & $D_\text{train}$ = $3.13 \times 10^{5}$ & $\left(1.68^{+1.5}_{-0.15}\right)\times 10^{10}$ & $1.002 \pm 0.055$ & $\left(3.79^{+1.1}_{-0.28}\right)\times 10^{-4}$ \\
        \bottomrule
    \end{tabular}
    \caption{Fitted scaling parameters for $Zgggg$ MLP-I surrogates trained with MSE loss.}
    \label{tab:paramfit_zgggg_MLP_MSE}
\end{table}

\begin{table}[H]
    \centering
    \renewcommand{\arraystretch}{1.3}
    \begin{tabular}{l@{\hskip 15pt} c@{\hskip 15pt} c@{\hskip 15pt} c@{\hskip 15pt} c}
        \toprule
        param.\ $X$ &  2nd param. & $X_c$ & $\alpha_X$ & $K_X$ \\
        \midrule
        
$D_\text{train}$ & $N_\text{LLoCa-Tr}$ = $6\times10^{6}$&  $3.76^{+0.44}_{-0.20}$ & $0.911 \pm 0.013$ & $\simeq 0$ \\
        \bottomrule
    \end{tabular}
    \caption{Fitted scaling parameters for $Zgggg$ LLoCa-Transformer surrogates trained with MSE loss.}
    \label{tab:paramfit_zgggg_lloca_MSE}
\end{table}

\begin{table}[H]
    \centering
    \renewcommand{\arraystretch}{1.3}
    \begin{tabular}{l@{\hskip 15pt} c@{\hskip 15pt} c@{\hskip 15pt} c@{\hskip 15pt} c}
        \toprule
        param.\ $X$ &  2nd param. & $X_c$ & $\alpha_X$ & $K_X$ \\
        \midrule
        \hline
\multirow{3}{*}{$D_\text{train}$}
 & $N_\text{MLP}$ = $10^{4}$ & $3.71^{+0.69}_{-0.24}$ & $1.588 \pm 0.028$ & $\simeq 0$ \\
 & $N_\text{MLP}$ = $10^{5}$ & $5.77^{+0.84}_{-0.34}$ & $1.738 \pm 0.027$ & $\simeq 0$ \\
 & $N_\text{MLP}$ = $10^{6}$ & $5.08^{+0.77}_{-0.31}$ & $1.720 \pm 0.027$ & $\simeq 0$ \\\hline
\multirow{3}{*}{$C$ [FLOPs]}
 & $D_\text{train}$ = $10^{2}$ & $\left(5.12^{+2.1}_{-0.41}\right)\times 10^{10}$ & $1.88 \pm 0.20$ & $\left(3.59^{+0.63}_{-0.23}\right)\times 10^{-3}$ \\
 & $D_\text{train}$ = $10^{4}$ & $\left(7.50^{+2.3}_{-0.57}\right)\times 10^{10}$ & $1.565 \pm 0.049$ & $\left(2.02^{+0.76}_{-0.16}\right)\times 10^{-6}$ \\
 & $D_\text{train}$ = $3.86 \times 10^{5}$ & $\left(11.58^{+1.9}_{-0.72}\right)\times 10^{10}$ & $1.812 \pm 0.030$ & $\left(2.02^{+11}_{-0.20}\right)\times 10^{-9}$ \\                   
        \bottomrule
    \end{tabular}
    \caption{Fitted scaling parameters for $WZ$ MLP-I surrogates trained with MSE loss.}
    \label{tab:paramfit_wz_MLP_MSE}
\end{table}

\begin{table}[H]
    \centering
    \renewcommand{\arraystretch}{1.3}
    \begin{tabular}{l@{\hskip 15pt} c@{\hskip 15pt} c@{\hskip 15pt} c@{\hskip 15pt} c}
        \toprule
        param.\ $X$ &  2nd param. & $X_c$ & $\alpha_X$ & $K_X$ \\
        \midrule
\hline
\multirow{3}{*}{$D_\text{train}$}
 & $N_\text{MLP}$ = $10^{4}$ & $14.58^{+3.0}_{-0.99}$ & $1.223 \pm 0.029$ & $\simeq 0$ \\
 & $N_\text{MLP}$ = $10^{5}$ & $\left(1.63^{+0.29}_{-0.10}\right)\times 10^{1}$ & $1.310 \pm 0.028$ & $\simeq 0$ \\
 & $N_\text{MLP}$ = $10^{6}$ & $13.21^{+2.7}_{-0.88}$ & $1.262 \pm 0.029$ & $\simeq 0$\\\hline
\multirow{3}{*}{$C$ [FLOPs]}
 & $D_\text{train}$ = $10^{2}$ & $\left(6.35^{+3.1}_{-0.53}\right)\times 10^{10}$ & $1.98 \pm 0.59$ & $\left(7.14^{+1.2}_{-0.44}\right)\times 10^{-2}$ \\
 & $D_\text{train}$ = $10^{4}$ & $\left(10.69^{+4.6}_{-0.87}\right)\times 10^{10}$ & $1.506 \pm 0.088$ & $\left(2.89^{+0.67}_{-0.20}\right)\times 10^{-4}$ \\
 & $D_\text{train}$ = $3.82 \times 10^{5}$ & $\left(7.74^{+2.7}_{-0.60}\right)\times 10^{10}$ & $1.307 \pm 0.036$ & $\left(1.39^{+1.61}_{-0.13}\right)\times 10^{-6}$ \\
        \bottomrule
    \end{tabular}
    \caption{Fitted scaling parameters for $WZg$ MLP-I surrogates trained with MSE loss.}
    \label{tab:paramfit_wzg_MLP_MSE}
\end{table}

\begin{table}[H]
    \centering
    \renewcommand{\arraystretch}{1.3}
    \begin{tabular}{l@{\hskip 15pt} c@{\hskip 15pt} c@{\hskip 15pt} c@{\hskip 15pt} c}
        \toprule
        param.\ $X$ &  2nd param. & $X_c$ & $\alpha_X$ & $K_X$ \\
        \midrule
\hline
\multirow{4}{*}{$D_\text{train}$}
 & $N_\text{MLP}$ = $10^{2}$ & $\left(4.72^{+0.81}_{-0.30}\right)\times 10^{1}$ & $1.154 \pm 0.058$ & $\left(5.28^{+1.51}_{-0.39}\right)\times 10^{-3}$ \\
 & $N_\text{MLP}$ = $10^{4}$ & $\left(2.67^{+0.65}_{-0.19}\right)\times 10^{1}$ & $0.995 \pm 0.030$ & $\simeq 0$ \\
 & $N_\text{MLP}$ = $10^{5}$ & $\left(1.83^{+0.56}_{-0.14}\right)\times 10^{1}$ & $0.959 \pm 0.034$ & $\simeq 0$ \\
 & $N_\text{MLP}$ = $10^{6}$ & $\left(1.36^{+0.78}_{-0.12}\right)\times 10^{1}$ & $0.932 \pm 0.037$ & $\simeq 0$ \\\hline
\multirow{3}{*}{$C$ [FLOPs]}
 & $D_\text{train}$ = $10^{2}$ & $\left(1.12^{+7.0}_{-0.11}\right)\times 10^{10}$ & $1.61 \pm 1.0$ & $\left(2.12^{+0.31}_{-0.13}\right)\times 10^{-1}$ \\
 & $D_\text{train}$ = $10^{4}$ & $\left(2.66^{+1.5}_{-0.23}\right)\times 10^{10}$ & $1.39 \pm 0.12$ & $\left(2.39^{+0.42}_{-0.15}\right)\times 10^{-3}$ \\
 & $D_\text{train}$ = $3.88 \times 10^{5}$ & $\left(1.58^{+0.97}_{-0.14}\right)\times 10^{10}$ & $1.055 \pm 0.041$ & $\left(3.90^{+1.6}_{-0.31}\right)\times 10^{-5}$ \\
                   
        \bottomrule
    \end{tabular}
    \caption{Fitted scaling parameters for $WZgg$ MLP-I surrogates trained with MSE loss.}
    \label{tab:paramfit_wzgg_MLP_MSE}
\end{table}

\begin{table}[H]
    \centering
    \renewcommand{\arraystretch}{1.3}
    \begin{tabular}{l@{\hskip 15pt} c@{\hskip 15pt} c@{\hskip 15pt} c@{\hskip 15pt} c}
        \toprule
        param.\ $X$ &  2nd param. & $X_c$ & $\alpha_X$ & $K_X$ \\
        \midrule
        \hline
\multirow{3}{*}{$D_\text{train}$}
 & $N_\text{MLP}$ = $10^{4}$ & $13.7^{+2.7}_{-0.91}$ & $1.273 \pm 0.029$ & $\simeq 0$ \\
 & $N_\text{MLP}$ = $10^{5}$ & $10.6^{+2.3}_{-0.72}$ & $1.256 \pm 0.028$ & $\simeq 0$ \\
 & $N_\text{MLP}$ = $10^{6}$ & $8.43^{+2.0}_{-0.59}$ & $1.227 \pm 0.028$ & $\simeq 0$ \\\hline
\multirow{3}{*}{$C$ [FLOPs]}
 & $D_\text{train}$ = $10^{2}$ & $\left(5.14^{+4.5}_{-0.46}\right)\times 10^{10}$ & $1.79 \pm 0.48$ & $\left(4.99^{+0.82}_{-0.31}\right)\times 10^{-2}$ \\
 & $D_\text{train}$ = $10^{4}$ & $\left(7.95^{+4.5}_{-0.68}\right)\times 10^{10}$ & $1.405 \pm 0.083$ & $\left(2.29^{+0.56}_{-0.16}\right)\times 10^{-4}$ \\
 & $D_\text{train}$ = $3.85 \times 10^{5}$ & $\left(6.75^{+2.5}_{-0.53}\right)\times 10^{10}$ & $1.316 \pm 0.037$ & $\left(1.15^{+1.2}_{-0.11}\right)\times 10^{-6}$ \\                   
        \bottomrule
    \end{tabular}
    \caption{Fitted scaling parameters for $WWZ$ MLP-I surrogates trained with MSE loss.}
    \label{tab:paramfit_wwz_MLP_MSE}
\end{table}

\begin{table}[H]
    \centering
    \renewcommand{\arraystretch}{1.3}
    \begin{tabular}{l@{\hskip 15pt} c@{\hskip 15pt} c@{\hskip 15pt} c@{\hskip 15pt} c}
        \toprule
        param.\ $X$ &  2nd param. & $X_c$ & $\alpha_X$ & $K_X$ \\
        \midrule
        \hline
\multirow{3}{*}{$D_\text{train}$}
 & $N_\text{MLP}$ = $10^{4}$ & $14.1^{+2.6}_{-0.92}$ & $1.304 \pm 0.029$ & $\simeq 0$ \\
 & $N_\text{MLP}$ = $10^{5}$ & $\left(1.72^{+0.25}_{-0.10}\right)\times 10^{1}$ & $1.461 \pm 0.027$ & $\simeq 0$ \\
 & $N_\text{MLP}$ = $10^{6}$ & $14.7^{+2.2}_{-0.89}$ & $1.445 \pm 0.027$ & $\simeq 0$ \\\hline
\multirow{3}{*}{$C$ [FLOPs]}
 & $D_\text{train}$ = $10^{2}$ & $\left(6.17^{+7.4}_{-0.57}\right)\times 10^{10}$ & $1.21 \pm 0.28$ & $\left(7.98^{+1.5}_{-0.52}\right)\times 10^{-2}$ \\
 & $D_\text{train}$ = $10^{4}$ & $\left(1.74^{+0.43}_{-0.12}\right)\times 10^{11}$ & $1.681 \pm 0.079$ & $\left(1.42^{+0.34}_{-0.10}\right)\times 10^{-4}$ \\
 & $D_\text{train}$ = $4.44 \times 10^{5}$ & $\left(13.6^{+3.0}_{-0.93}\right)\times 10^{10}$ & $1.516 \pm 0.033$ & $\left(1.73^{+3.1}_{-0.16}\right)\times 10^{-7}$ \\
                   
        \bottomrule
    \end{tabular}
    \caption{Fitted scaling parameters for $\gamma\gamma g$ MLP-I surrogates trained with MSE loss.}
    \label{tab:paramfit_aag_MLP_MSE}
\end{table}

\begin{table}[H]
    \centering
    \renewcommand{\arraystretch}{1.3}
    \begin{tabular}{l@{\hskip 15pt} c@{\hskip 15pt} c@{\hskip 15pt} c@{\hskip 15pt} c}
        \toprule
        param.\ $X$ &  2nd param. & $X_c$ & $\alpha_X$ & $K_X$ \\
        \midrule
        \hline
\multirow{3}{*}{$D_\text{train}$}
 & $N_\text{MLP}$ = $10^{4}$ & $\left(1.49^{+0.50}_{-0.12}\right)\times 10^{1}$ & $0.939 \pm 0.031$ & $\simeq 0$ \\
 & $N_\text{MLP}$ = $10^{5}$ & $11.77^{+3.8}_{-0.90}$ & $0.988 \pm 0.029$ & $\simeq 0$ \\
 & $N_\text{MLP}$ = $10^{6}$ & $9.50^{+3.4}_{-0.74}$ & $0.967 \pm 0.029$ & $\simeq 0$ \\
\hline
\multirow{3}{*}{$C$ [FLOPs]}
 & $D_\text{train}$ = $10^{2}$ & $\left(1.35^{+2.7}_{-0.13}\right)\times 10^{10}$ & $1.03 \pm 0.27$ & $\left(12.04^{+2.0}_{-0.75}\right)\times 10^{-2}$ \\
 & $D_\text{train}$ = $10^{4}$ & $\left(4.45^{+1.8}_{-0.36}\right)\times 10^{10}$ & $1.261 \pm 0.083$ & $\left(1.79^{+0.36}_{-0.12}\right)\times 10^{-3}$ \\
 & $D_\text{train}$ = $4.44 \times 10^{5}$ & $\left(3.66^{+1.3}_{-0.29}\right)\times 10^{10}$ & $1.112 \pm 0.034$ & $\left(1.60^{+0.99}_{-0.14}\right)\times 10^{-5}$ \\
        \bottomrule
    \end{tabular}
    \caption{Fitted scaling parameters for $\gamma\gamma gg$ MLP-I surrogates trained with MSE loss.}
    \label{tab:paramfit_aagg_MLP_MSE}
\end{table}

\clearpage
\bibliography{tilman,refs_1,refs_2}
\end{document}

%% file: incl_settings.tex
\usepackage[utf8]{inputenc} 
\usepackage[T1]{fontenc} 	
\usepackage[english]{babel} 

\usepackage[bitstream-charter]{mathdesign}
\usepackage{geometry} 		
\usepackage{amsmath} 		
\usepackage{mathtools} 		
\usepackage{float} 			
\usepackage{graphicx} 		
\usepackage{tabularx} 		
\usepackage{booktabs} 		
\usepackage{color, xcolor} 	
\usepackage{pdfpages} 		
\usepackage{extarrows} 		
\usepackage{multirow} 		
\usepackage{multicol} 		
\usepackage{enumitem} 		
\usepackage{xspace} 		
\usepackage{stackrel} 		
\usepackage{tikz} 			
\usepackage{braket} 		
\usepackage{bm} 			
\usepackage{tensor} 		
\usepackage{slashed} 		
\usepackage{siunitx} 		
\usepackage{lastpage} 		
\usepackage{cite} 			
\usepackage[normalem]{ulem} 
\usepackage{fontawesome} 	
\usepackage{tocloft} 		
\usepackage{titlesec} 		
\usepackage{doi} 			
\usepackage{hyperref} 		
\usepackage[most]{tcolorbox} 					
\usepackage[nameinlink, capitalize]{cleveref} 	
\usepackage[nottoc, notlot, notlof]{tocbibind} 	
\usepackage[ruled, vlined]{algorithm2e} 		
\usepackage{makecell}
\usepackage[makeroom]{cancel}
\usepackage{feynmf}
\usepackage{cancel}
\usepackage{nicefrac}

\makeatletter
\def\BState{\State\hskip-\ALG@thistlm}
\makeatother

\makeatletter
\@ifundefined{pdfoutput}{}{\DeclareGraphicsRule{*}{mps}{*}{}}
\makeatother

\makeatletter
\DeclareRobustCommand*{\bfseries}{%
   \not@math@alphabet\bfseries\mathbf
   \fontseries\bfdefault\selectfont
   \boldmath
}
\makeatother

\hypersetup{
	pdftitle={Scaling laws for amplitude surrogates},
	pdfauthor={},
	colorlinks=true, 			
	linkcolor={red!50!black}, 	
	citecolor={blue!50!black}, 	
	urlcolor={blue!80!black} 	
} 

\DeclareSymbolFont{usualmathcal}{OMS}{cmsy}{m}{n}
\DeclareSymbolFontAlphabet{\mathcal}{usualmathcal}

\SetArgSty{textnormal}
\SetKwComment{Comment}{{\small\#}~}{}
\SetCommentSty{mycommfont}

\setitemize{itemsep=0pt, parsep=0pt} 				
\setenumerate{itemsep=0pt, parsep=0pt} 				
\setitemize{itemsep=2pt,topsep=2pt,parsep=0pt,partopsep=0pt,leftmargin=*}
\setenumerate{itemsep=0pt,topsep=2pt,parsep=0pt,partopsep=0pt,labelindent=3pt,leftmargin=*}
\newlist{todolist}{itemize}{2}
\setlist[todolist]{label=$\square$}
\usepackage{pifont}

\usepackage{amsmath}
 
\usepackage{amsthm} 		
\theoremstyle{definition}

\usetikzlibrary{arrows, shapes.geometric, arrows.meta, shapes, decorations.pathreplacing, fit, patterns, patterns.meta}

\definecolor{Rcolor}{HTML}{E99595}
\definecolor{Gcolor}{HTML}{C5E0B4}
\definecolor{Bcolor}{HTML}{9DC3E6}
\definecolor{Ycolor}{HTML}{FFE699}
\definecolor{Ycolor_light}{HTML}{FFF7DE}
\definecolor{Gcolor_light}{HTML}{F1F8ED}

\tikzstyle{expr} = [circle, minimum width=1.8cm, minimum height=1.8cm, text centered, align=center, inner sep=0, draw,font=\LARGE]
\tikzstyle{txt_huge} = [align=center, font=\Huge, scale=2]
\tikzstyle{txt} = [align=center, font=\LARGE, minimum height=1cm]
\tikzstyle{cinn} = [double arrow, double arrow head extend=0cm, double arrow tip angle=130, shape border rotate=90, inner sep=0, align=center, minimum width=2.1cm, minimum height=2.3cm, fill=Bcolor, draw,font=\LARGE]
\tikzstyle{cinn_black} = [cinn, minimum height=2.5cm, fill=black]
\tikzstyle{arrow} = [thick,-{Latex[scale=1.0]}, line width=0.2mm, color=black]
\tikzstyle{loss} = [rectangle, align=center,  minimum width=1.8cm, minimum height=1.5cm,fill=Rcolor,font=\LARGE, rounded corners]
\tikzstyle{xt} = [rectangle, align=center,  minimum width=5cm, minimum height=1.5cm,fill=Gcolor,font=\LARGE, rounded corners]
\tikzstyle{xts} = [rectangle, align=center,  minimum width=1cm, minimum height=1.5cm,fill=Gcolor,font=\Large, rounded corners]

\tikzstyle{embed} = [rectangle, rounded corners=0.3ex, minimum width=1.5cm, minimum height=1cm, text centered, align=center, inner sep=0, fill=Ycolor, font=\large, draw]

\tikzstyle{small_cinn} = [double arrow, double arrow head extend=0cm, double arrow tip angle=130, inner sep=0, align=center, minimum width=1.1cm, minimum height=0.5cm, fill=Bcolor, draw]

\tikzstyle{transformer} = [rectangle, rounded corners, minimum width=6cm, minimum height=2.4cm, font=\large, fill=Gcolor_light, draw]

\tikzstyle{attention} = [rectangle, rounded corners=0.3ex, minimum width=5.5cm, minimum height=1.2cm, align=center, fill=Gcolor, draw, font=\large]

%% file: incl_shortcuts.tex
\definecolor{red_cb}{HTML}{e41a1c}
\definecolor{blue_cb}{HTML}{377eb8}
\definecolor{green_cb}{HTML}{4daf4a}
\definecolor{purple_cb}{HTML}{984ea3}
\definecolor{orange_cb}{HTML}{ff7f00}

\definecolor{EmeraldGreen}{HTML}{1ea78d}
\definecolor{EnglishRed}{HTML}{b02427}
\hypersetup{colorlinks=true,urlcolor=EmeraldGreen,citecolor=EmeraldGreen,linkcolor=EnglishRed}

\newcommand{\XLangle}{\Bigl\langle}
\newcommand{\XRangle}{\Bigr\rangle}

\newcommand\one{\leavevmode\hbox{\small1\normalsize\kern-.33em1}}

\newcommand{\normal}{\mathcal{N}} 			

\newcommand{\mean}[1]{\left\langle#1\right\rangle}

\newcommand{\loss}{\mathcal{L}} 	

\newcommand{\arXiv}[2][]{%
	\ifthenelse{\equal{#1}{}}%
	{\href{http://arxiv.org/abs/#2}{arXiv:#2}}%
	{\href{http://arxiv.org/abs/#2}{arXiv:#2~[#1]}}}

\def\slashchar#1{\setbox0=\hbox{$#1$}           
   \dimen0=\wd0                                 
   \setbox1=\hbox{/} \dimen1=\wd1               
   \ifdim\dimen0>\dimen1                        
      \rlap{\hbox to \dimen0{\hfil/\hfil}}      
      #1                                        
   \else                                        
      \rlap{\hbox to \dimen1{\hfil$#1$\hfil}}   
      /                                         
   \fi}

\newcommand{\tikznode}[2]{%
\ifmmode%
\tikz[remember picture,baseline=(#1.base),inner sep=0pt] \node (#1) {$#2$};%
\else
\tikz[remember picture,baseline=(#1.base),inner sep=0pt] \node (#1) {#2};%
\fi}

\def\mathswitchr#1{\relax\ifmmode{\text{#1}}\else$\text{#1}$\xspace\fi}
\def\mathswitch#1{\relax\ifmmode#1\else$#1$\xspace\fi}